\documentclass[a4paper,twocolumn,10pt,accepted=2022-01-19]{quantumarticle}
\pdfoutput=1

\usepackage{float}
\floatstyle{boxed}
\usepackage{wrapfig}
\usepackage{array}
\usepackage{graphicx}
\usepackage{amsbsy}
\usepackage[utf8x]{inputenc}
\usepackage{epstopdf}
\usepackage{amsmath,amssymb,amsfonts}
\usepackage{dsfont}
\usepackage{cprotect}

\usepackage[sort&compress,numbers,square,comma]{natbib}
\usepackage[dvipsnames]{xcolor}

\usepackage{enumitem}

\newcommand{\bra}[1]{\langle #1|}
\newcommand{\ket}[1]{|#1\rangle}
\newcommand{\braket}[1]{\left\langle #1 \right\rangle}

\newcommand{\expec}[1]{\left\langle #1 \right\rangle}

\newcommand{\crea}[1]{\hat{a}^\dagger_{#1}}
\newcommand{\ann}[1]{\hat{a}_{#1}}

\renewcommand{\eqref}[1]{\mbox{Eq.~(\ref{#1})}}
\renewcommand{\Re}[1]{{\rm Re}\left(#1 \right)}
\renewcommand{\Im}[1]{{\rm Im}\left(#1 \right)}

\newcommand{\be}{\begin{equation}}
\newcommand{\ee}{\end{equation}}
\newcommand{\bea}{\begin{eqnarray}}
\newcommand{\eea}{\end{eqnarray}}

\usepackage{url}
\usepackage[colorlinks]{hyperref}
\hypersetup{
	plainpages=true,
	breaklinks=true,
	hypertexnames=false,
	pageanchor=true,
	colorlinks=true,
	linkcolor={blue},
	citecolor={red},
	urlcolor={blue},
	anchorcolor={black}
}

\newcommand{\LL}{\mathcal{L}}
\newcommand{\EE}{\mathcal{E}}
\newcommand{\FF}{\mathcal{F}}
\newcommand{\DD}{\mathcal{D}}
\newcommand{\rhot}{\hat{\rho}(t)}

\newcommand{\sss}{\hat{\rho}_{\rm ss}}
\newcommand{\eig}[1]{\hat{\rho}_{#1}}

\makeatletter
\newcommand*\bigcdot{\mathpalette\bigcdot@{.5}}
\newcommand*\bigcdot@[2]{\mathbin{\vcenter{\hbox{\scalebox{#2}{$\m@th#1\bullet$}}}}}
\makeatother

\usepackage[normalem]{ulem} 

\usepackage{algorithm}
\usepackage{algpseudocode}

\begin{document}

	\author{Fabrizio Minganti}
	\email{fabrizio.minganti@gmail.com} 
	\affiliation{Institute of Physics, Ecole Polytechnique Fédérale de Lausanne (EPFL), CH-1015 Lausanne, Switzerland}
	\affiliation{Center for Quantum Science and Engineering, Ecole Polytechnique F\'ed\'erale de Lausanne (EPFL), CH-1015 Lausanne, Switzerland}
	\orcid{0000-0003-4850-1130}
	\author{Dolf Huybrechts}
	\email{dolf.huyb@gmail.com} 
	\affiliation{TQC, Universiteit Antwerpen, Universiteitsplein 1, B-2610 Antwerpen, Belgium}
	\affiliation{Univ Lyon, Ens de Lyon, CNRS, Laboratoire de Physique, F-69342 Lyon, France}
    \thanks{\\ The authors contributed equally}
	\orcid{0000-0002-5821-3493}
	\date{8/02/2022}

	\title{Arnoldi-Lindblad time evolution: Faster-than-the-clock algorithm for the spectrum of time-independent and Floquet open quantum systems}
	
	\begin{abstract}
	
	The characterization of open quantum systems is a central and recurring problem for the development of quantum technologies. 
	For time-independent systems, an (often unique) steady state describes the average physics once all the transient processes have faded out, but interesting quantum properties can emerge at intermediate timescales.
    Given a Lindblad master equation, these properties are encoded in the spectrum of the Liouvillian whose diagonalization, however, is a challenge even for small-size quantum systems.
    Here, we propose a new method to efficiently provide the Liouvillian spectral decomposition. We call this method an Arnoldi-Lindblad time evolution, because it exploits the algebraic properties of the Liouvillian superoperator to efficiently construct a basis for the Arnoldi iteration problem.
	The advantage of our method is double: 
	(i) It provides a faster-than-the-clock method to efficiently obtain the steady state, meaning that it produces the steady state through time evolution shorter than needed for the system to reach stationarity. 
	(ii) It retrieves the low-lying spectral properties of the Liouvillian with a minimal overhead, allowing to determine both which quantum properties emerge and for how long they can be observed in a system.
	This method is \textit{general and model-independent}, and lends itself to the study of large systems where the determination of the Liouvillian spectrum can be numerically demanding but the time evolution of the density matrix is still doable. 
	Our results can be extended to time evolution with a time-dependent Liouvillian. 
	In particular, our method works for Floquet (i.e., periodically driven) systems, where it allows 
	not only to construct the Floquet map for the slow-decaying processes, but also to retrieve the stroboscopic steady state and the eigenspectrum of the Floquet map.
    Although the method can be applied to any Lindbladian evolution (spin, fermions, bosons, \dots), for the sake of simplicity we demonstrate the efficiency of our method on several examples of coupled bosonic resonators (as a particular example). Our method outperforms other diagonalization techniques and retrieves the Liouvillian low-lying spectrum even for system sizes for which it would be impossible to perform exact diagonalization.
	\end{abstract}

	\maketitle

\section{Introduction}

The high-degree of controllability of photonic platforms, such as superconducting circuits \cite{YouNat11, SchoelkopfNat08, CarmichaelPRX15, FinkPRX17, FitzpatrickPRX17}, Rydberg atoms \cite{Mueller_2012,BernienNAT2017}, and optomechanical resonators \cite{AspelmeyerRMP14,Teufel2011, Kolkowitz1603, GilSantosPRL17}, makes them ideal candidates for the realization of quantum simulators and quantum computers \cite{AruteNat19, SchneiderOptica19, AndersenNatPhys20}. 
This fostered intense experimental research on the control of the quantum properties of these systems, ranging from the creation of (ultra)strong photon-matter interaction \cite{Carusotto_RMP_2013_quantum_fluids_light,kockum2019} to the engineering of the environment to generate properties unachievable in closed systems \cite{VerstraeteNATPH2009, SorientePRR21, DicandiaArxiv21}.
Simultaneously, much theoretical effort was dedicated to the determination of an open system's properties. In particular, understanding  the properties of the steady state, reached after long enough time evolution,
has been of central interest \cite{KesslerPRA12,BucaNPJ2012,AlbertPRA14, MingantPRA18_Spectral,NigroJSM19}. 
In open quantum systems the interplay between classical and quantum fluctuations induced by the competition between Hamiltonian dynamics, pumping, and dissipation results in a nonequilibrium stationary state whose properties cannot be determined by a free energy analysis \cite{Gardiner_BOOK_Quantum,Haroche_BOOK_Quantum,DaleyAdvancesinPhysics2014}.

Many efficient numerical methods have been developed in the last years, each one with its advantages and drawbacks, to find the steady state of systems described by a Lindblad master equation (i.e., for a weakly interacting and Markovian environment).
To cite some, Ref. \cite{CaoPRA16} proposes a variational ansatz valid for two weakly-coupled resonators. 
Methods based on mean-field approximations determine efficiently the state of lattices of nonlinear resonators \cite{LeBoitePRL13,WeimerPRL15,JinPRX16,BiondiPRA17, CasteelsPRA18, HuybrechtsPRA19}, but their predictions are ensured to be valid only for weak nonlinearities or for high-connectivity models \cite{HuybrechtsPRB20}.
Monte Carlo \cite{NagyPRA18}, tensor network \cite{CuiPRL15,MascarenasPRA15,KshetrimayumNAT17}, and neural network \cite{NagyPRL19,VicentiniPRL19,HartmannPRL19} algorithms proved efficient even for highly-entangled phases, but can be burdened by various factors, such as the exponential growth of a density matrix, entropy, or correlations. 
Methods exploiting the permutational symmetry reduce the computational complexity work for Dicke-like models \cite{GeggSciRep17,ShammahPRA18}, but cannot be applied to planar architectures characterizing, e.g., superconducting circuits lattices. 

Although the steady state is fundamental to correctly characterize an open quantum system, many relevant quantum properties emerge in the system's transient dynamics. For instance, highly-entangled states or quantum properties can often emerge in a well-initialized system \cite{VerstraeteNATPH2009,LeghtasScience15,JinPRX16,PurinpjQI17}. Correctly understanding the states and the timescales for which these properties can be witnessed is an even more challenging quest.
Some pioneering techniques attempted to efficiently study the dynamics of open quantum systems.
For example, renormalization procedures \cite{FinazziPRL15,RotaPRB17} to reduce the size of the Hilbert space, truncated Wigner approximations \cite{Foss-FeigPRA17,VicentiniPRA18}, semi-classical approaches \cite{LeBoitePRL13,CasteelsPRA16,CasteelsPRA17-2}, Gaussian ansatzes \cite{VerstraelenPRA20}, (Gutzwiller \cite{HuybrechtsPRA20}) Monte Carlo quantum trajectory methods \cite{BartoloEPJST17,RotaNJP18,SanchezPRA19}, and matrix product states \cite{HartmannPRL2010}, all managed to study the dynamics of an open system under certain approximations or assumptions.

In principle, one can determine the steady state and all the dynamical features of an open quantum system by diagonalizing the so-called Liouvillian superoperator, that is, the linear function which induces the time evolution of a system's density matrix via the Lindblad master equation \cite{LidarLectureNotes}. 
However, the Liouvillian diagonalization is extremely costly and inefficient due to the quartic scaling of the Liouvillian superoperator with the already exponentially large dimension of the system's Hilbert space.
This ``bad scaling'' makes it almost impossible to compute the spectrum of even few-site problems.
Furthermore, iterative methods are known to require preconditioning to converge to physically meaningful results \cite{NationArXiv15}. 
The time evolution of the density matrix with the Lindblad master equation is relatively more feasible thanks to its ``merely'' quadratic scaling with the system's Hilbert space dimension.
However, without the Liouvillian eigendecomposition, it is impossible to know \textit{a priori} how a certain initial state evolves in time, and to do that one would need to evolve any initial state with the Lindblad master equation. 

The determination of the Liouvillian spectrum also plays a fundamental role for dissipative critical phenomena--such as (boundary) time crystals \cite{IeminiPRL18,SeiboldPRA20,MingantiarXiv20}, dissipative phase transitions \cite{KesslerPRA12,MingantPRA18_Spectral}, and more exotic effects emerging such as dissipative freezing \cite{SanchezPRA19} and synchronization \cite{TindallNJP20}--where the nontrivial role of the Liouvillian long-lived metastable states allows to correctly determine the scaling towards the thermodynamic limit.
These systems are candidate for quantum information and metrology, with perspectives for quantum technologies \cite{MacieszczakPRA16,RotaPRL19,DicandiaArxiv21,Munozarxiv20}.
As such, a correct determination of the effect of errors and their possible corrections is a focal problem also in quantum information and quantum error correction \cite{AlbertPRA18,LieuPRL20}.
Similarly, the Liouvillian spectrum is fundamental to correctly describe Liouvillian exceptional points (EPs) \cite{MingantiPRA19,HatanoMP19,MingantiPRA2020,ArkhipovPRA20, ArkhipovPRA21,KumarArxi21,KumarArxi21_2}, an emerging field where new experimental results open perspectives to the study of larger systems with high-number of modes \cite{NaghilooNatPhys19,ChenArxiv21}.

Furthermore, there are classes of (driven) open quantum systems where the Liouvillian operator is time-dependent itself. Within this paradigm, a particularly relevant class is that of Floquet (or periodically driven) systems \cite{GongPRL18, Lazarides_PhysRevRes20, Sato_2020, Ikeda_SciAdv20, ikeda2021nonequilibrium, Restrepo_PRL2016}, where an external element induces a periodic modulation of the Hamiltonian and/ or of the dissipators. For this class of systems, a steady state may never be reached. Nevertheless, by stroboscopically checking the system at integer periods, the density matrix reaches a stroboscopic steady state. In this case, the properties of the system are captured by the so-called Floquet map, that is, the evolution operator of one period. 
As such, the spectrum of the Floquet map contains information about possible emergence of multiple stroboscopic steady states and the creation of Floquet time crystals \cite{Hartmann_NJP2017}. The diagonalization of a Floquet map of a dissipative system, however, is even more difficult than the corresponding diagonalization of the Liouvillian superoperator, since it requires to explicitly build such a map even before dealing with the determination of the eigenstates and eigenvalues.

In the following, we will introduce a new method capable of combining the exactness of the Liouvillian and Floquet map diagonalization with the efficiency of time evolution of $\rhot$ to retrieve extremely precise estimations of the (stroboscopic) steady state and of the long-lived dynamical processes from relatively short time evolution of the density matrix.
To do that, we will use the information encoded in the time evolution to efficiently reconstruct the spectrum of the Liouvillian or of the Floquet map in the time independent or dependent cases, respectively.
In particular, in the latter case, our method will also efficiently construct the Floquet map for the slow-decaying processes.
\textit{In other words, our method combines the advantages of both the time evolution methods and those of the diagonalization, being only partially affected by their disadvantages.}
Thus, our method allows determining the spectral properties of systems where the density matrix dynamics can be (approximately) determined, but $\LL$ cannot  be diagonalized. An implementation of this method in Python can be found in Ref.~\cite{arnoldilindbladgithub}.

The structure of the article is the following.
In Sec.~\ref{Sec:Evolutions_standard} we introduce the time-independent Lindblad master equation and the associated Liouvillian superoperator.
Sec.~\ref{Sec:Methods} contains the Arnoldi-Lindblad time evolution algorithm.
At first, we justify our idea from a physical perspective (Sec.~\ref{Sec:General_idea}).
In Sec.~\ref{Sec:Krylov_definition}, we introduce the mathematical objects used to define the Arnoldi-Lindblad time evolution algorithm in Sec.~\ref{Sec:Arnoldi-Lindblad}. 
Examples of the efficiency and validity of the algorithm are provided in Sec.~\ref{Sec:casestudy}. 
In Sec.~\ref{Sec:floquet} we then extend our algorithm to the study of Floquet Lindblad master equation, showing its efficiency in determining the Floquet map for the slow-evolving eigenvalues and eigenmatrices.
Conclusions and Perspectives are drawn in Sec.~\ref{Sec:Conclusion}.
The appendices contain the pseudocodes (App.~\ref{App:Pseudo-codes}) and a discussion of the failure of standard iterative methods for the determination of the Liouvillian spectrum (App.~\ref{App:Iterative}).

\section{Time-independent Lindblad master equation}
\label{Sec:Evolutions_standard}
We begin by considering the dynamics of a time-independent open quantum system, the Floquet time-dependent case being considered in Sec.~\ref{Sec:floquet}. 
Accordingly, the system dynamics is encoded in the Lindblad master equation ($\hbar=1$)\cite{BreuerBookOpen, Haroche_BOOK_Quantum}
\begin{equation}\label{Eq:LME}
    \partial_t \rhot = -i \left[ \hat{H}, \rhot \right] + \sum_{\mu} \DD \left[ \hat{J}_{\mu} \right] \rhot,
\end{equation}
where $\rhot$ is the system density matrix and $\hat{H}$ is the Hamiltonian describing the coherent evolution of the system. $\DD \left[ \hat{J}_{\mu} \right]$ are Lindblad dissipators associated with the jump operator $\hat{J}_{\mu} $, and they read
\begin{equation}\label{Eq:Dissipator}
    \DD \left[ \hat{J}_{\mu} \right] = \hat{J}_{\mu}\rhot \hat{J}_{\mu}^\dagger -\frac{\hat{J}_{\mu}^\dagger \hat{J}_{\mu} \rhot +\rhot \hat{J}_{\mu}^\dagger \hat{J}_{\mu} }{2}.
\end{equation}
Each Lindblad master equation admits (at least) one steady state $\sss$ \cite{RivasBOOK_Open}, i.e., that state which does not evolve any more under the action of the Lindblad master equation
\begin{equation}\label{Eq:Steady_state}
    \partial_t \sss = 0.
\end{equation}

It can be shown that if $\sss$ is unique, any initial state converges to it, meaning that $\sss$ represents the long-time dynamics of the system.
Thus, for systems with a unique steady state, the most straightforward way to find $\sss$ is to evolve an initial state $\hat{\rho}(0)$ to a time $t$ sufficiently long, so that $\rhot \simeq \sss$.
In the presence of slow processes, or in the case one is also interested in determining the characteristics of specific transient processes, this method can turn out to be quite inefficient.


\subsection{Liouvillian spectrum}

The Lindblad master equation can be recast in terms of the so-called Liouvillian superoperator $\LL$ \cite{LidarLectureNotes}, i.e.,
\begin{equation}
    \partial_t \rhot = \LL \rhot.
\end{equation}
We call the Liouvillian a superoperator because it acts on operators to produce new operators. 
We will indicate, except if stated differently, operators with a hat (e.g., $\hat{O}$) and superoperators with calligraphic symbols (e.g., $\mathcal{O}$).
$\LL$ allows to solve \eqref{Eq:LME} and to express the evolution of any initial state as
\begin{equation}\label{Eq:evolution}
    \rhot=\exp(\LL t) \hat{\rho}(0).
\end{equation} 

Similarly to Hamiltonian problems, where the diagonalization of $\hat{H}$ allows to express the dynamics of the system in terms of the energies and eigenvectors,
by diagonalizing $\LL$ one obtains the eigenvalues $\lambda_j$ and eigenmatrices (eigenstates) $\eig{j}$ such that
\begin{equation}\label{Eq:eigen}
    \LL \eig{j} = \lambda_j \eig{j}.
\end{equation}
While the eigenvalues represent the timescales of the system, the $\eig{j}$ describe the states explored along the system dynamics. 
We order the eigenvalues by their real part, so that $|\operatorname{Re}(\lambda_0)|\leq|\operatorname{Re}(\lambda_1)|\leq|\operatorname{Re}(\lambda_2)|\dots$ 
All the eigenvalues have negative real part [$\operatorname{Re}(\lambda_j)\leq 0$].
Within such a representation, $\sss \propto \eig{0}$ and $\lambda_0=0$. Therefore, the diagonalization of the Liouvillian immediately retrieves the steady state as one of its eigenmatrices.
The Liouvillian spectrum, and in particular its low-lying part, turns out to be quite useful in several problems, ranging from phase transitions to the determination of long-lasting quantum properties \cite{MingantPRA18_Spectral,MacieszczakPRL16, KrimerPRL19,SeiboldPRA20,MingantiarXiv20,LieuPRL20}.

To numerically study the properties of a superoperator with the standard tools of numerical linear algebra, one can recast $\LL$ it in its matrix form. 
Indeed, since the Liouvillian is linear [i.e., $\LL (\alpha \hat{\sigma}+ \beta \hat{\chi}) = \alpha  \LL \hat{\sigma}+ \beta \LL\hat{\chi}$ for any operator $\hat{\sigma}$ and $\hat{\chi}$, and for any complex number $\alpha$ and $\beta$], it can be represented as a matrix.
This is equivalent to write the operators as a column vector (a procedure called vectorization): a matrix $N\times N$ becomes a vector of size $N^2$ where the consecutive elements of a row (row-major) or the consecutive elements of a column (column-major) reside next to each other\footnote{\label{footnote:1}To vectorize an operator means, for example, to transform the matrix 
$$\hat{A} = \begin{pmatrix} a & b \\ c & d \end{pmatrix}$$ into the vector $$\vec{A} = \begin{pmatrix} a \\ b \\ c \\ d \end{pmatrix}$$ Consequently, superoperators become matrices. For a more detailed discussion, see, e.g., \cite{MingantiPHD,MingantiPRA19}.}.
Therefore, diagonalizing the Liouvillian in its matrix form is extremely challenging, because $\LL$ is a matrix $N^2 \times N^2$.
Methods which allow to efficiently determine the spectrum of $\LL$ without encountering these scaling problems are extremely useful \cite{NakagawaPRL21,PopkovPRL21}. 

\begin{figure*}
    \centering
    \includegraphics[width=0.98 \textwidth]{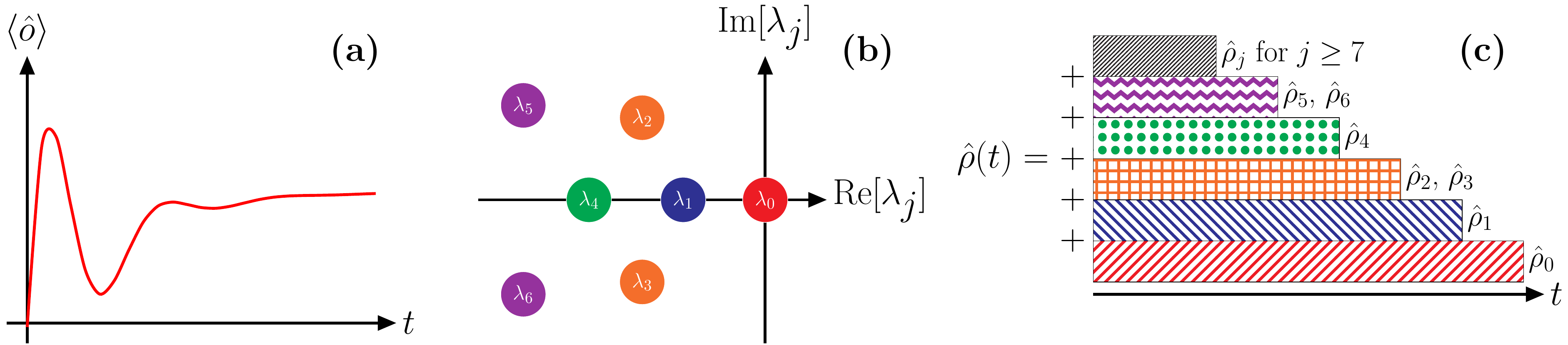}
    \caption{Pictorial representation of the physical meaning of the time evolution of a density matrix from a spectral point of view. (a) The dynamics of a system governed by a Lindblad master equation has always some common characteristics. Indeed, after an initial transient dynamics, the system converges to its steady state. (b) These characteristics are clear from a Liouvillian spectrum point of view. Except from the steady state $\eig{0}$ characterized by $\lambda_0=0$, the eigenvalues characterizing the Liouvillian eigenmatrices have always non-positive real part (they describe decaying processes towards the steady-state) and they always appear as complex conjugate (they preserve the Hermiticity of a density matrix). (c) This translates in a specific form of the density matrix along its evolution. While at the beginning all the eigenmatrices are relevant, as time passes the coefficients of the eigenmatrices $\eig{j}$ whose eigenvalues $\lambda_j$ have large negative part become negligible. But while at long time $\eig{j}$ with large $j$ are irrelevant, the slow-decaying eigenmatrices (including the steady state) are present also in the short-time dynamics. Thus, by just doing long-time evolution, we are discarding all the possible information on $\eig{j}$ derived from the short-time dynamics.}
    \label{fig:spectrum_extraction_alg}
\end{figure*}

\section{Arnoldi time evolution}
\label{Sec:Methods}

In any open quantum system, dissipation introduces a typical timescale $\tau = 1/ \Gamma$, where $\Gamma$ is a dissipation rate.
For instance, in optical resonators, such a timescale can be represented by the photon lifetime.
In this regard, one can roughly distinguish two ``families'' of $\eig{j}$: those that decay quickly (short-lived states, i.e., $|\text{Re}(\lambda_j)| > \Gamma$) and those that decay slowly (long-lived states, i.e., $|\text{Re}(\lambda_j)| <\Gamma$). 
The short-lived states describe the initial and quick dynamics of an open quantum system, making it difficult to exploit them for technological advantage.
Thus, one is often interested in \textit{characterizing only the long-lived states}. 
\textit{Our algorithm accomplishes this task without the Liouvillian diagonalization complexity}.

\subsection{General idea behind the algorithm and its physical meaning}
\label{Sec:General_idea}

Knowing the eigendecomposition of the Liouvillian (i.e. its eigenvalues and eigenvectors) it is possible to determine the time evolution of an open quantum system (see Fig.~\ref{fig:spectrum_extraction_alg}), because
any initial state can be recast as \cite{LidarLectureNotes}
\begin{equation}\label{Eq:eigendecomposition}
    \hat{\rho}(t=0)=\sss + \sum_{j\geq 1} c_j \eig{j}.
\end{equation}
From \eqref{Eq:eigendecomposition}, and using Eqs.~(\ref{Eq:evolution})~and~(\ref{Eq:eigen}), we have 
\begin{equation}\label{eq:spectralsol}
    \rhot = \sss+ \sum_{j\geq 1} c_j e^{\lambda_j t} \eig{j}.
\end{equation}
In other words, it is trivial to see that, if we know $\eig{j}$ and $\lambda_j$, we can determine the time evolution of any density matrix by applying appropriate coefficients.
The other way around it is also possible.
Since all the $\Re{\lambda_j}\leq 0$ \cite{RivasBOOK_Open,BreuerBookOpen}, after a sufficient long time the system reaches its steady state and $\rhot \simeq \sss$.
Knowing $\eig{0}$ (and thus $\sss$), we can proceed ``backwards'' in time and determine all the other eigenmatrices. For instance, if we know $\eig{0}$, and $\rhot = \sss + c_1(t) \eig{1}$, we can determine $\eig{1}$ by extrapolation [see also the sketch in Fig.~\ref{fig:spectrum_extraction_alg} (c)].

This determination of the spectrum from the time dynamics is very inefficient and does not bring any numerical advantage to the determination of $\sss$.
However, by determining $\sss$ with a long time evolution, we are ``throwing away'' all the information about the Liouvillian spectrum accumulated along the dynamics [see again Fig.~\ref{fig:spectrum_extraction_alg} (c)].
To improve the method, we notice that at sufficiently long time, the physics of the system is confined in the manifold spanned by the eigenmatrices $\eig{j}$ that are the slowest decaying.
Indeed, the dynamics is described by those $M$ eigenmatrices such that $\lambda_j/ \Gamma$ is sufficiently close to zero (we recall $\tau= 1/\Gamma$ is the typical decaying timescale of the system), i.e.,
\begin{equation}\label{Eq:approximated_rho}
    \hat{\rho}(t> \tau) \simeq \sss+ \sum_{1\leq j < M} c_j e^{\lambda_j t} \eig{j}.
\end{equation}
To correctly describe $\hat{\rho}( t> \tau)$, we do not need to exactly know the $M$ eigenmatrices $\eig{j}$, but just a basis of this $M$-dimensional manifold. 
We can project the Liouvillian onto this reduced basis, obtaining an \textit{effective} Liouvillian much smaller than the original one, reducing the computational cost of the diagonalization.

The advantages of this method are evident. Not only it does shorten the time needed to obtain the steady state, but it also makes possible to obtain an excellent estimation of the slow-decaying eigenmatrices of the Liouvillian.
Having clarified the physical idea behind the algorithm, the mathematical formalization of this intuition can be provided in terms of Krylov subspaces, and the proposed algorithm will be a reinterpretation of the time evolution of an open quantum system in terms of Arnoldi iteration. Thus, we call such a procedure an \textit{Arnoldi-Lindblad time evolution}.


\subsection{Arnoldi iteration and the evolution operator}
\label{Sec:Krylov_definition}

Although a detailed description of Krylov subspaces and Arnoldi iteration goes beyond the purpose of this article (we refer the interested reader to, e.g., Ref.~\cite{TrefethenBOOK}), here we briefly discuss them. We will use the notation of operators and superoperators for consistency with the following discussion on the Arnoldi-Lindblad method.

\subsubsection{Krylov subspaces and Arnoldi iteration}

Suppose we want to determine part of the spectrum of a superoperator, say the Liouvillian $\LL$.
This task can be achieved via an iterative diagonalization known as Arnoldi iteration. The key idea behind this algorithm is that one can recursively apply $n$ times the Liouvillian to a random matrix $\hat{\sigma}$ of norm one, producing the so-called Krylov matrix
\begin{equation}\label{Eq:Krylov}
    K_n = \left\{\hat{\sigma}, \LL \hat{\sigma}, \LL^2 \hat{\sigma} \dots , \LL^n \hat{\sigma}  \right\}.
\end{equation}
This method will highlight those eigenmatrices $\eig{j} $ associated with the $\lambda_j$ of \textit{largest} absolute value. Indeed, using \eqref{Eq:eigendecomposition}, $\hat{\sigma}$ becomes
\begin{equation}\label{Eq:Krylov_large}
    \hat{\sigma}=\sum_i c_i \eig{i} \Rightarrow \LL^n \hat{\sigma}=\sum_i c_i \lambda_i^{n}\eig{i},
\end{equation}
since, by definition, $\LL \eig{i}=\lambda_i \eig{i}$.

\begin{figure*}[t!]
    \centering
    \includegraphics[width=0.85\textwidth]{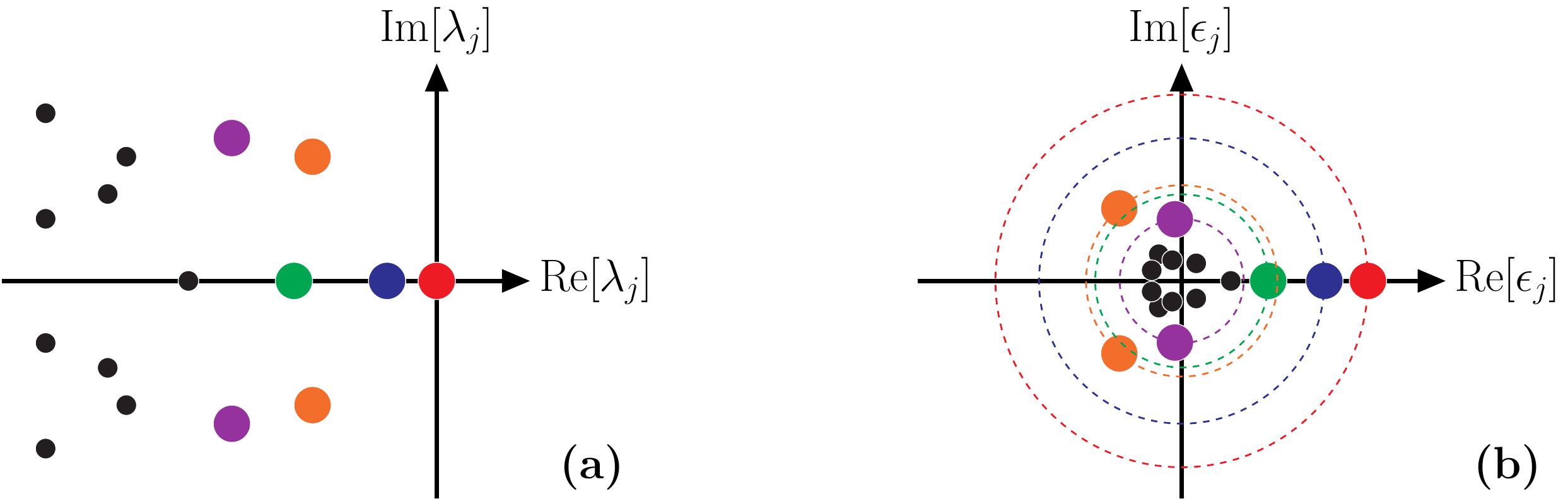}
    \caption{Spectrum of the Liovillian [dots in (a)] vs spectrum of the corresponding evolution operator [dots in (b) with the same color]. While the steady state is the zero of the Liouvillian (red dot), $\epsilon_0$ is the largest eigenvalue of $\EE$, as it can be seen in (b), where the dashed lines represent the circles of radius $|\epsilon_j|=\exp[\operatorname{Re}(\lambda_j)]$. Thus, the Arnoldi iteration is well suited to obtain $\eig{0}$ from $\EE$. Furthermore, the eigenvalues with the smallest real part acquire the larger magnitude [other colored dots in (a) and (b)], while the fast-decaying processes are condensed towards the zero (black dots).}
    \label{fig:transformation_spectrum}
\end{figure*}

The Arnoldi iteration builds up an orthonormal basis from $K_n$ in \eqref{Eq:Krylov} via Gram-Schmidt orthonormalization.
If one calls $\hat{\sigma}_{1}\dots\hat{\sigma}_{n}$ the Arnoldi orthonormal basis obtained from $K_n$, and $\mathcal{S}_n$ the rectangular matrices whose columns are the $n$ vectorized Arnoldi operators constituting the Arnoldi basis (see footnote \textsuperscript{\ref{footnote:1}}), one can define an effective Liouvillian as
\begin{equation}\label{Eq:Definition_effective_Liouvillian}
    \LL^{\rm eff}_n = \mathcal{S}_n^{\dagger} \LL \mathcal{S}_n.
\end{equation}
In the limit in which $n\to N^2$, where $N$ is the dimension of the Hilbert space, $\LL^{\rm eff}$ becomes the full Liouvillian $\LL$, $\mathcal{S}$ being nothing but a change of basis.

The diagonalization of $\LL^{\rm eff}$ for $n\ll N^2$ is, in principle, efficient: $\LL^{\rm eff}$ is a relatively small and upper Hessenberg matrix (i.e., ``almost'' triangular):
\begin{equation}\label{Eq:Effective_Liouvillian}
    \mathcal{L}^{\rm eff}_n =\begin{bmatrix}
   l_{1,1} & l_{1,2} & l_{1,3} & \cdots  & l_{1,n} \\
   l_{2,1} & l_{2,2} & l_{2,3} & \cdots  & l_{2,n} \\
   0       & l_{3,2} & l_{3,3} & \cdots  & l_{3,n} \\
   \vdots  & \ddots  & \ddots  & \ddots  & \vdots  \\
   0       & \cdots  & 0     & l_{n,n-1} & l_{n,n}
\end{bmatrix}.
\end{equation}
Importantly, one can diagonalize $\LL^{\rm eff}_n$ and obtain the (Ritz) eigenvalues $\lambda_{\rm j}^{\rm eff}(n)$ and eigenmatrices $\eig{j}^{\rm eff}(n)$.
In the limit in which the $\eig{j}^{\rm eff}(n)$ approximate all the wanted $\eig{j}$,
one can stop the iteration.

Notably, the coefficients $l_{i,j}$ are directly computed by the algorithm and one never needs to explicitly build the change-of-basis matrix $\mathcal{S}_n$ in \eqref{Eq:Definition_effective_Liouvillian} to obtain $\mathcal{L}^{\rm eff}_n$.
The pseudocode for the Arnoldi iteration is provided in Algorithm~\ref{Alg:ArnoldiIteration} in App.~\ref{App:Pseudo-codes}.

\subsubsection{The exponential map as an alternative operator}

There are several reasons for which directly computing the Liouvillian spectrum via the Arnoldi iteration is impractical for large systems.
First, it is necessary to store the Liouvillian matrix elements, which can be costly.
Second, the Krylov subpaces tends to bring out the Liouvillian eigenvalues with the largest magnitude [c.f. \eqref{Eq:Krylov_large}].
These problems can be mitigated by using the shift-and-inverted Arnoldi method. 
In this way, the most relevant vectors become those closest to zero \cite{TrefethenBOOK}.
This, however, can either increase the numerical cost of the diagonalization if exact LU decomposition is performed on $\LL$, or can lead to significant numerical errors due to instabilities if solved iteratively by the application of $\hat{H}$ and $\hat{J}_\mu$
\cite{NationArXiv15}.
We provide a brief discussion of these issues in Appendix~\ref{App:Iterative}.

Let us consider, instead, the evolution superoperator
\begin{equation}\label{Eq:Evolution_operator}
    \EE=\exp(\LL T),
\end{equation} where $T$ is a generic time.
The first remark is that there is a direct correspondence between the eigenvalues of $\EE$ and those of $\LL$. Indeed, one has
\begin{equation}\label{Eq:correspondence}
    \EE \eig{j} = \epsilon_j \eig{j}= e^{\lambda_j T} \eig{j} \quad \Leftrightarrow \quad \LL \eig{j}=\lambda_j \eig{j}.
\end{equation}
The second fundamental remark is that the Liouvillian is the generator of a contraction semigroup \cite{RivasBOOK_Open}.
Equivalently, since $\lambda_0=0$ and Re$(\lambda_j)\leq 0$, $\epsilon_0=e^{\lambda_0 T}=1$ is the \textit{largest} eigenvalue of $\EE$.
Similarly, the $\lambda_j$ with the real part closest to zero will become the largest $\epsilon_j$.
This can be seen in Fig.~\ref{fig:transformation_spectrum}, where we show how the spectrum of $\LL$ transforms into that of $\EE$.
The Krylov matrix for the operator $\mathcal{E}$ will therefore enhance the slow-decaying processes, and it will read
\begin{equation}\begin{split}
    K_n &= \left\{\hat{\rho}(0),\, \EE \hat{\rho}(0),\, \EE^2 \hat{\rho}(0)\, \dots ,\, \EE^n \hat{\rho}(0)  \right\}.
\end{split}
\end{equation}
Using Arnoldi iteration, we can therefore construct
\begin{equation}\label{Eq:Effective_Evolution}
    \EE^{\rm eff}_n= \mathcal{S}^\dagger_n\EE \mathcal{S}_n =\begin{bmatrix}
   e_{1,1} & e_{1,2} & e_{1,3} & \cdots  & e_{1,n} \\
   e_{2,1} & e_{2,2} & e_{2,3} & \cdots  & e_{2,n} \\
   0       & e_{3,2} & e_{3,3} & \cdots  & e_{3,n} \\
   \vdots  & \ddots  & \ddots  & \ddots  & \vdots  \\
   0       & \cdots  & 0     & e_{n,n-1} & e_{n,n}
\end{bmatrix},
\end{equation}
where this time $\mathcal{S}_n$ are the rectangular matrices whose columns are the $n$ vectorized Arnoldi basis operators of the Krylov matrix of the evolution superoperator.
Although we have avoided the problems of the shift-and-inverted Arnoldi method, we encounter a far greater numerical cost.
Indeed, the computation of the exponential of the Liouvillian is a numerically very costly procedure, comparable to the initial full-diagonalization problem.
Furthermore, we are still dealing with superoperators, making the treatment of the problem challenging for large systems.

\subsection{Krylov space of the time evolution and the Arnoldi-Lindblad algorithm}
\label{Sec:Arnoldi-Lindblad}

Let us analyze more in detail the time evolution of an open system.
Even if we never write the Liouvillian, or its exponential, we can always formally write
\eqref{Eq:evolution}.
If we now introduce $T=t/N$, we have
\begin{equation}\label{Eq:Evolution_operator_N}
    \rhot = \left[\exp{(\LL T)}\right]^N \hat{\rho } (0)=\mathcal{E}^N \hat{\rho } (0).
\end{equation} 
Within this representation, if we stock the density matrix evolved with a Lindblad master equation at each integer multiple of $T$, we obtain
\begin{equation}\begin{split}
    K_n &= \left\{\hat{\rho}(0),\, \hat{\rho}(T),\, \hat{\rho}(2T),\, \dots\, \hat{\rho}(nT) \right\}\\
     &=\left\{\hat{\rho}(0),\, \EE \hat{\rho}(0),\, \EE^2 \hat{\rho}(0)\, \dots ,\, \EE^n \hat{\rho}(0)  \right\}.
\end{split}
\end{equation}
Therefore, we can extract the Krylov matrix of $\mathcal{E}$ from the dynamics of $\rhot$ without ever writing $\mathcal{L}$ or $\EE$.
Having $K_n $, we can apply the Arnoldi iteration to determine the spectrum of $\EE$.
And knowing the eigenstates of $\EE$, we can use \eqref{Eq:correspondence} to directly compute the spectrum of $\LL$.
This procedure has roughly the same numerical cost as a simple time evolution.

The last missing ingredient is how to determine $T$. 
Although this part can be heavily optimized (a task we plan to undergo in the future), a very simple idea comes from deciding a priori in which part of the spectrum one is interested.
Similarly to \eqref{Eq:approximated_rho}, if one wants to know the part of the spectrum such that $|\Re{\lambda_j}|<\Gamma$, then it would make sense to consider a $T$ which is a fraction of $(1/\Gamma)$.
This procedure works even in cases where the system undergoes critical phenomena associated with the emergence of few small eigenvalues, e.g., first-order dissipative phase transitions or second-order ones with breaking of $\mathbb{Z}_N$ symmetries \cite{MingantPRA18_Spectral}.

Summing up, we have introduced a technique that, by just computing the time evolution of one state according to the Lindblad master equation, allows obtaining the slow-decaying part of the evolution map.
This is an efficient method because it allows to determine the Liouvillian eigendecomposition even for large system, where the full diagonalization would be impossible.
Such a reduced Liouvillian map, can then be applied to any arbitrary initial state, allowing to easily determine the time dynamics once fast-decaying processes (not described by the reduced Liouvillian) play virtually no role.
Notice that, due to the approximated nature of $\LL_n^{\rm eff}$, the evolution map is no more completely-positive and describes a physical system at short time only when considering a well-defined superposition of \textit{only} those eigenmatrices which reached convergence in the algorithm. Equivalently, the reduced Liouvillian cannot be used to describe the physics at short times, since it could provide nonphysical results. For numerical reasons we evolve the initial state up to time $T$ and then orthonormalize it with respect to the previous elements of the Arnoldi basis. this time-evolution orthonormalization procedure, together with the contruction of the effective Liouvillian superoperator allows efficiently determining the Liouvillian eigendecomposition. We call this modified time evolution an Arnoldi-Lindblad master equation.

Notice also that, the simple recipe presented above may not be working for systems near dissipative critical points where the typical decay time of many modes may become significantly smaller than $1/\Gamma$, e.g., systems entering a Zeno regime where many modes are very fast and many become slower than $(1/\Gamma)$ \cite{rosso2021arxiv1,RossiniPRA2021,rosso2021arxiv2}. 
In such specific cases, the very fact that there is a proliferation of ``slow'' processes makes it impossible to reduce the number of basis eigenvectors below a certain number, because the algorithm by its own nature selects the states such that $|\Re{1/\lambda_j}|>T$. 
This problem is inherent to the structure of the diagonalization problem and would affect any ``extrapolation'' method.
Nonetheless, in some cases (e.g., Refs.~\cite{MingantiarXiv20,mingantiArxvi21}), symmetries play a fundamental role in the emergence of these many slow states, and the system dynamics can be separated into independent symmetry sectors, each one containing only a limited number of slow states. 
In these cases, the problem of the proliferation of slow states can be mitigated by choosing  an appropriate initial state $\hat{\rho}(0)$ belonging just to one symmetry sector: since $\EE$ will never mix states with different symmetry, the Arnoldi-Lindblad algorithm will preserve the symmetry along the dynamics, thus making it possible to separately determine the slow eigenvalues of each symmetry sector.  
A different and general approach would be to increase $T$ (thus making the eigenvalues $\epsilon_j$ more distant one from the other), but this would imply an increased numerical cost in the time-evolution step.

Furthermore, the algorithm's convergence depends on the initial state. Indeed, selecting a state with greater overlap with respect to the wanted eigenstates will ensure a faster convergence.
However, due to renormalization at each time step of the Arnoldi iteration and given that the Lindblad master equation quickly eliminates fast-decaying vectors, a suboptimal choice of $\hat\rho(0)$ does not affect very much the algorithm (indeed, in all the examples below, we checked that the number of iterations is almost constant regardless of the randomly chosen initial state).

Having clarified the limits of validity of this algorithm and its interest, we plan in the future to implement a variation of the presented algorithm employing, e.g., inflation and deflation techniques that at each time step provide a way to better select the converging eigenvalues and eliminate those with bad convergence \cite{TrefethenBOOK}, ensuring an ever wider range of applicability and higher efficiency.

The pseudocode to realize the Arnoldi-Lindblad time evolution and determine the spectrum of the low-lying part of the Liouvillian is detailed in Algorithm~\ref{Alg:ArnoldiTimeEvolution} in App.~\ref{App:Pseudo-codes}.
Comparing this procedure with the intuitive idea in Fig.~\ref{fig:spectrum_extraction_alg}, the steady state obtained from a long time dynamics corresponds to taking only $\EE^M \hat{\rho}(0)$ with very large $M$, while discarding all the information stored in the construction of the Krylov subspace $K_n$.
At this large time, $\EE^M \hat{\rho}(0)$ has a dominant projection over the steady state and many small contributions from the other slow eigenmodes.
This is the mathematical formalization of the intuition depicted in Fig.~\ref{fig:spectrum_extraction_alg}(c).

In what follows, we will always consider exact time evolution of the density matrix via Runge-Kutta integration of the equations of motion.
In these systems, we will consider sizes where $\hat{H}$ can be diagonalized on our machines, but $\LL$ cannot (thus having a ``rule of thumb'' to know if we can perform the system time evolution).
However, our method works independently on how the various $\hat{\rho}(nT)$ have been obtained, and it can be readily applied to compute the Liouvillian spectrum.
Furthermore, when considering extended problems, the size of the effective Liouvillian tends to be independent of the system dimension.
Indeed, the low-lying part of the Liouvillian is only marginally affected by increasing the system size. 
Although critical phenomena can occur, these are represented by a few eigenvalues, and the large majority of the new Liouvillian eigenvalues obtained by increasing the system size will be characterized by a large (in modulus) real part.

\section{The time-independent driven-dissipative Bose-Hubbard model}
\label{Sec:casestudy}

To prove the efficiency of our method, we consider the driven-dissipative Bose-Hubbard model (DDBH).
We make this choice because the DDBH allows describing simultaneously several nontrivial effects (interaction-induced oscillations, the competition between drive and dissipation, etc\dots). Nevertheless, as it also stems from the general discussion above, our method can be applied to a large class of Liouvillian problems.

Consider a resonator, whose energy is $\omega$. Two photons interact with a Kerr-type nonlinearity of intensity $U$. By joining two or more resonators together (e.g., by evanescent wave coupling in micropillars \cite{Carusotto_RMP_2013_quantum_fluids_light} or via capacitive or inductive coupling in superconducting circuits \cite{Gu2017}), an effective hopping term $J$ emerges, describing the passage of a photon from one resonator to the other.
The model is driven by a coherent laser of intensity $F_l (\crea{l} e^{i \omega_p t}+ e^{-i \omega_p t} \ann{l})$ ($l$ indicating the site, and $\omega_p$ being the pump frequency), where the operators $\crea{l}$ ($\ann{l}$) are the creation (annihilation) operators of the $l$-th site. 
The DDBH describes such an array of $L$ resonators.

Passing from the density matrix in the laboratory frame $\hat{\rho}_{\rm lab}(t)$ to the one in the frame rotating at the pump frequency via the transformation $\rhot = \hat{R}^\dagger \rhot  \hat{R}$, where $\hat{R}=\prod_l \exp(i \omega_p \crea{l}\ann{l} t)$, and introducing the pump-to-cavity detuning $\Delta= \omega_p - \omega$, the Hamiltonian reads
\begin{equation}\label{Eq:DDBH_Hamiltonian}
\begin{split}
    \hat{H}&= \sum_{l=1}^{L} \left[ -\Delta \crea{l}\ann{l} + \frac{U}{2}\left(\crea{l}\right)^2  \left(\ann{l}\right)^2 + F_j \left(\crea{l}+\ann{l} \right) \right] \\
& \quad \quad  -\frac{J}{z} \sum_{\langle l,m \rangle}  \crea{l} \ann{m},
\end{split}
\end{equation}
where $\langle l,m \rangle$ indicates the sum on the nearest neighbours and $z$ is the coordination number.
All the following analysis will be performed considering the time-independent Hamiltonian in \eqref{Eq:DDBH_Hamiltonian}.
Notice that all observables of the form $\expec{\crea{l}\ann{m}}$ are identical in both reference frames.

In this time-independent example, we will consider a dissipation that acts locally and uniformly in each site at a rate $\gamma$ by ejecting single photons from each resonator. Thus, the Lindblad master equation and the associated Liouvillian read [c.f. Eqs. (\ref{Eq:LME}) and (\ref{Eq:Dissipator})]
\begin{equation}\label{Eq:DDBH_Liouvillian}
    \partial_t \rhot = \LL \rhot = -i \left[\hat{H}, \rhot \right] + \gamma \sum_{l=1}^{L} \DD[\ann{l}] \rhot.
\end{equation}

The single cavity problem ($L=1$) is the standard Kerr resonator, and has been analytically solved for its steady state \cite{Drummond_JPA_80_bistability,StannigelNJP12,RobertsPRX20}, while lattice-like models have been investigated through a variety of other methods \cite{LeBoitePRL13,CasteelsPRA17-2,CasteelsPRA17,VicentiniPRA18,Foss-FeigPRA17,BiondiPRA17,RodriguezPRL17,BartoloPRA16, HuybrechtsPRA20, VerstraelenPRA20, VerstraelenPREV20}, including the exact diagonalization (ED) of the full Liouvillian \cite{MingantPRA18_Spectral,CasteelsPRA17-2}.
The DDBH is known to be characterized by a dissipative phase transition in the thermodynamic limit of infinite cavities.
Furthermore, the emergence of time-crystal phases in asymmetrically driven cavities has been discussed in \cite{SeiboldPRA20,LLedoNJP20}.
As such, the finite-size DDBH provides an ideal benchmark for our method, combining the difficulty of an emerging criticality with the large size of the Hilbert space of the coupled resonators.

To keep the simulation unbiased, all the results below have been obtained using the QuTiP library \cite{qutip1,qutip2} both for time evolution and Liouvillian diagonalization.

\subsection{Uniform drive}\label{sec:uniformdrive}

In the following, we will consider the dimer and trimer cases (i.e., the number of sites $L$ is either 2 or 3)
We will start by considering two and three identical cavities, where the parameters will be fixed at $\Delta=5 \gamma$, $F_1=F_2 =F_3 =F=4.5 \gamma$, $U=20\gamma$, and $J/z = 10 \gamma$ (the hopping term is renormalized by the connectivity $z$, i.e., the number of nearest neighbours).

\subsubsection{The dimer}
\label{Sec:Dimer}
For $L=2$ in Eqs.~(\ref{Eq:DDBH_Hamiltonian}) and (\ref{Eq:DDBH_Liouvillian}), and for the parameters considered, we find that the part of the spectrum shown below has reached convergence for a cutoff of $n_{max}=7$, that is, we suppose that any element of a density matrix $\braket{p|\rhot|q}$ is zero if $p>n_{max}$ or $q>n_{max}$, where $\ket{p}$ and $\ket{q}$ are Fock states.
It follows that $\hat{H}$ is a $64 \times 64$ matrix, while the whole Liouvillian has size $4096 \times 4096$, a still diagonalizable object (both exactly and iteratively using the shift-invert Arnoldi method for the smallest eigenvalues).
We will test our algorithm on two types of tasks: (i) Determine just the steady state; (ii) Determine the $m+1$ slowest eigenvalues and eigenmatrices, noted $\hat{\rho}^{(m)}$. For instance, $\hat{\rho}^{(4)}$ means we have determined ($\lambda_0$, $\lambda_1$, \dots, $\lambda_4$) and ($\eig{0}$, $\eig{1}$, \dots, $\eig{4}$) up to convergence. Note that the complex conjugates of the obtained eigenvalues and eigenmatrices are implicitly taken into account, i.e., without additional computational cost the determination of e.g. $\hat{\rho}^{(4)}$ in certain cases can determine $\hat{\rho}^{(5)}$.  
We check if the eigenvalues converge every 10 timesteps. The condition of convergence we require is $\|\EE \eig{j} - \epsilon_j^{\rm eff}\eig{j}\|<10^{-3}$. Note however 
that there exist multiple possible indicators of convergence, e.g.,
$\left\vert\left \|\operatorname{Tr} \left[ \left(\eig{j}\right)^\dagger \LL   \eig{j} \right]  \right \|^2 - \left\|\operatorname{Tr} \left[ \left(\eig{j}\right)^\dagger \LL^2   \eig{j} \right]  \right \|\right\vert$.

\begin{figure*}[t]
    \centering
    \includegraphics[width=0.98\textwidth]{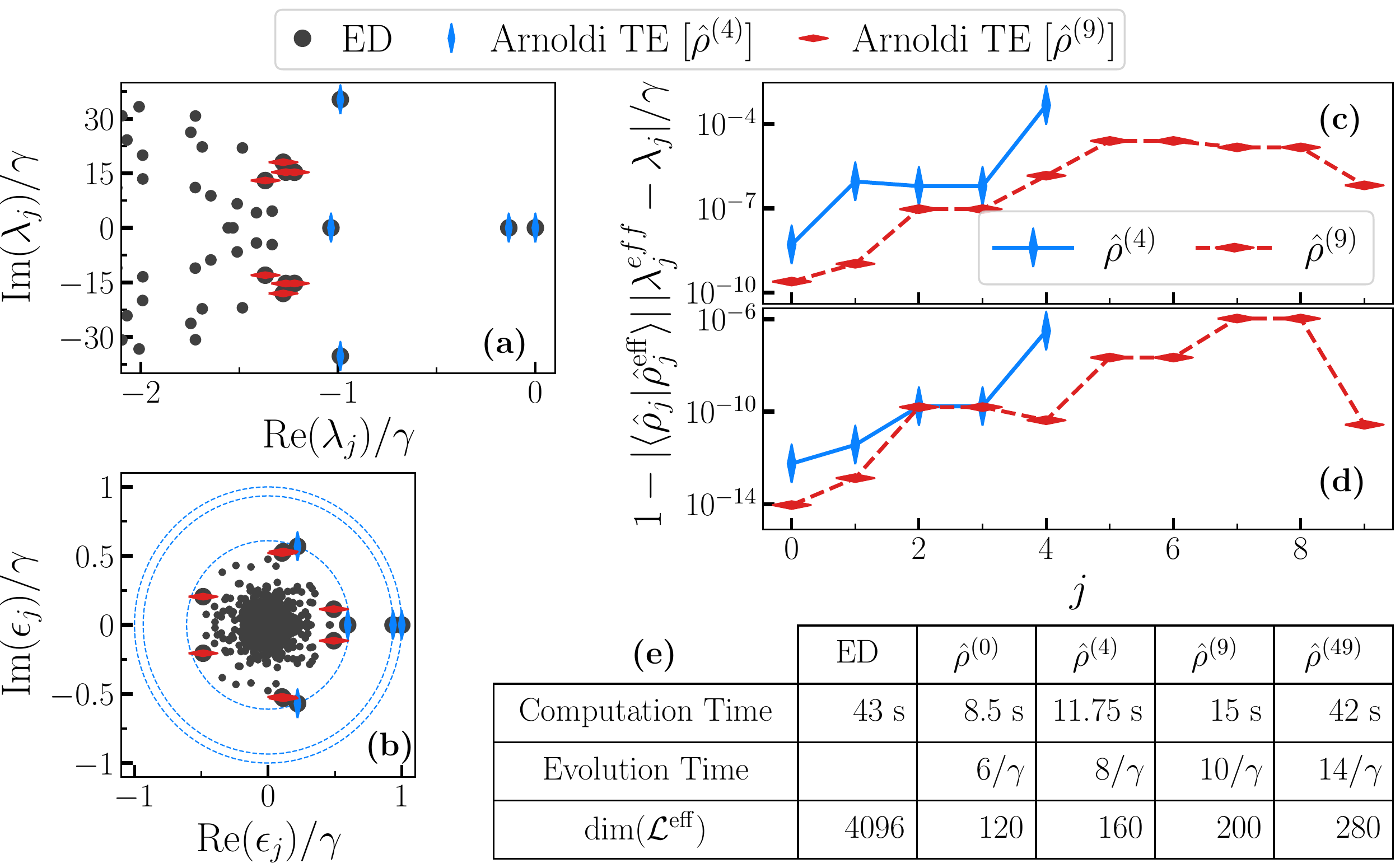}
    \caption{Bose Hubbard dimer for the parameters $\Delta = 5\gamma$, $F_1=F_2=4.5\gamma$, $U=20\gamma$, and $J = 10\gamma$, a cutoff $n_{max} = 7$ and $T=\gamma/20$. 
    The results are for: Exact diagonalization (ED, gray dots); our method, the Arnoldi-Lindblad time evolution for the 5 slowest states $\hat{\rho}^{(4)}$ (blue curves and vertical markers) and for the 10 slowest states $\hat{\rho}^{(9)}$ (red curves and horizontal markers).
    \textbf{(a)} Comparison of the Liouvillian spectrum obtained via ED and our method. 
    \textbf{(b)} Spectrum of the evolution operator obtained via ED and our method. The dashed lines represent the circles whose radius is the absolute value of $\epsilon_j$ (c.f. Fig.~\ref{fig:transformation_spectrum}). For the sake of graphical clarity, in this panel we considered $T=\gamma/2$ in order to make the eigenvalues more distant.
    \textbf{(c)} Difference between the eigenvalues obtained with the ED $\lambda_j$ and our method $\lambda_j^{\rm eff}$ for $\hat{\rho}^{(4)}$ and $\hat{\rho}^{(9)}$. \textbf{(d)} One minus the absolute value of the overlap between the eigenmatrices obtained via the ED and our method for $\hat{\rho}^{(4)}$ and $\hat{\rho}^{(9)}$.
    \textbf{(e)} Overview of the computational effort required for each task on our local machine [Intel(R) Xeon(R) W-2135 CPU @ 3.70GHz (12 CPUs) and 128 Gb of RAM].}
    \label{fig:dimer_spectrum_comparison}
\end{figure*}

\begin{figure}
    \centering
    \includegraphics[width=0.49\textwidth]{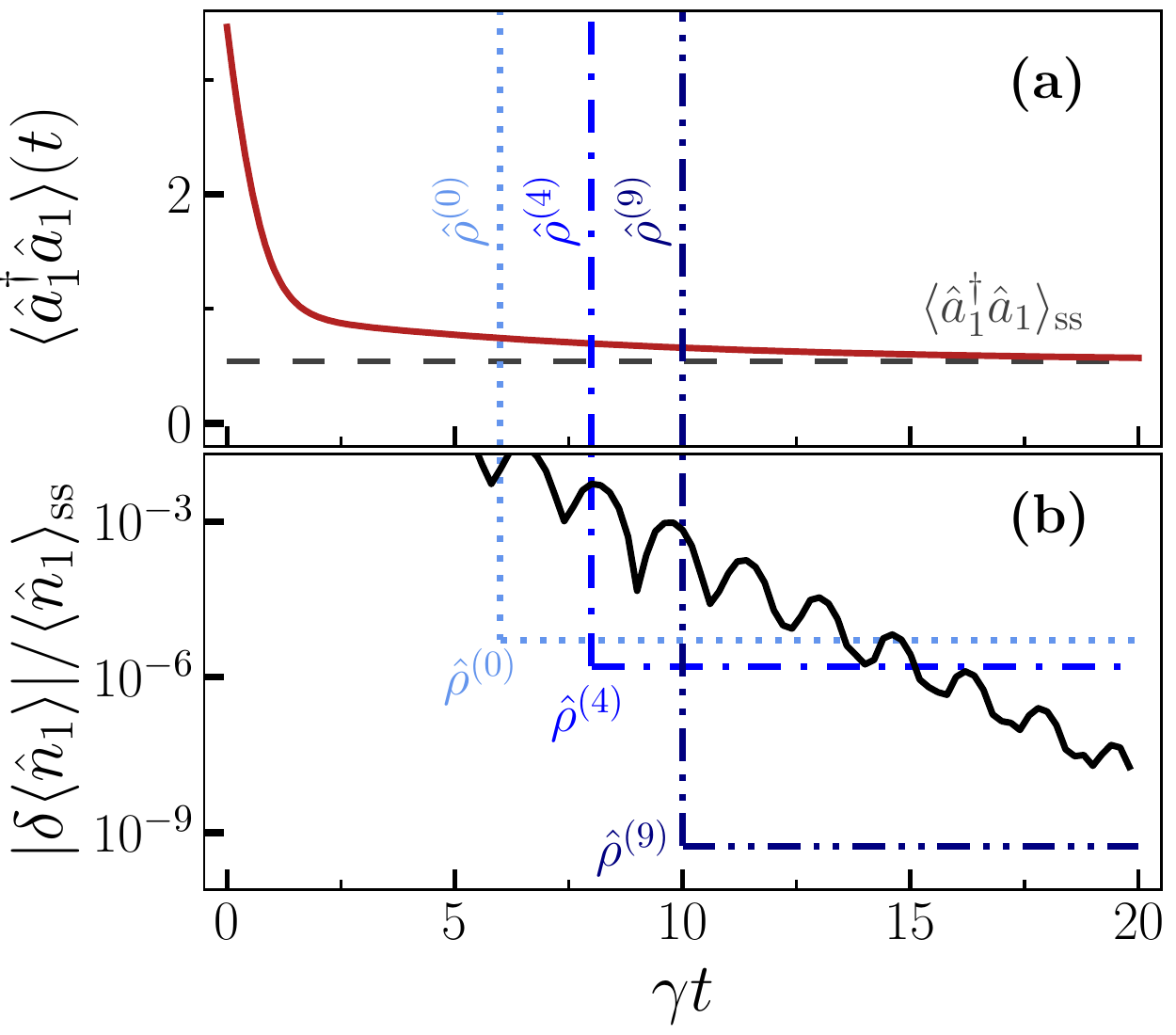}
    \caption{\textbf{(a)} Time evolution of the expectation value of the particle number of the first cavity of the dimer $\hat{n}_1 = \langle\hat{a}_1^\dagger\hat{a}_1\rangle$ (red full line). The steady-state solution obtained via exact diagonalization, i.e. $\langle\hat{a}_1^\dagger\hat{a}_1\rangle_{\rm ss}$ (gray horizontal dashed line). The time when our method reaches convergence for $\hat{\rho}^{(m)}$ is shown by vertical lines for $m=0$ (light-blue dotted line), $m=4$ (blue dash-dotted line) and $m=9$ (dark-blue dashed-dotted-dotted line). \textbf{(b)} Relative difference between the expectation value $\langle\hat{n}_1\rangle$ obtained via ED and either our method or a fit of the exponential decrease towards the steady state (black solid line).
    The fit of $\langle\hat{n}_1\rangle_{\rm ss}$ via \eqref{Eq:fit_curve} has been done with the \texttt{curve\char`_fit} function of the SciPy library \cite{SciPy}, and 
    the curve plotted here is a rolling average over a time $0.08 \Gamma$.
    Same parameters and initial state as in Fig.~\ref{fig:dimer_spectrum_comparison}.
    }
    \label{fig:particlenumber}
\end{figure}

\begin{figure}
    \centering
    \includegraphics[width=0.49 \textwidth]{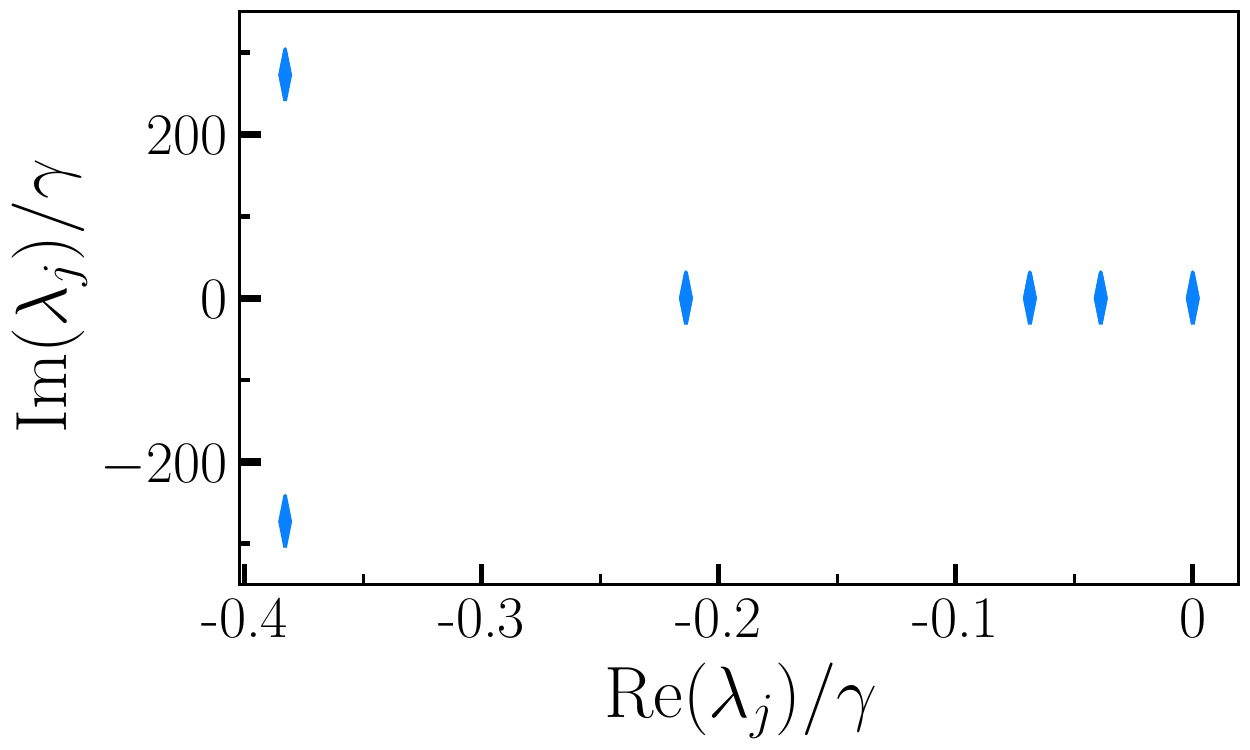}
    \caption{Results for $\hat{\rho}^{(5)}$ of the Bose Hubbard trimer for the parameters $\Delta = 5\gamma$, $F_1=F_2=4.5\gamma$, $U=20\gamma$, and $J = 10\gamma$. This is the same set of parameters as in Fig.~\ref{fig:dimer_spectrum_comparison}, having opportunely rescaled $J$ by the connectivity $z$. The results were obtained for the convergent cutoff $n_{max} = 7$. 
    These results were obtained after a computation time of less than 1 hour on our local machine.
    }
    \label{fig:trimer}
\end{figure}

An overview of the computational effort required for each task on our local machine is given in Fig.~\ref{fig:dimer_spectrum_comparison}(e).
Clearly, to obtain few values, for this limited system size, our method is much faster than a Liouvillian ED.
The shift-inverted Arnoldi method (using LU decomposition) to obtain the $5$ smallest eigenvalues required a time shorter than ED by a few seconds.
Nevertheless, if one is interested in getting many eigenvalues and eigenmatrices, our algorithms starts to become comparable with the Liouvillian ED [c.f. $\hat{\rho}^{(49)}$].

Besides the speed of the method, evidently one is also interested in the accuracy of its predictions.
In the other panels of Fig.~\ref{fig:dimer_spectrum_comparison} we compare the results obtained for $\hat{\rho}^{(4)}$ (blue vertical markers and lines) and $\hat{\rho}^{(9)}$ (red horizontal markers and lines) through our method with the corresponding ones obtained via ED. In Figs.~\ref{fig:dimer_spectrum_comparison}(a) and (b) we show the Liouvillian spectrum and the spectrum of the evolution operator, respectively.
There is an excellent agreement between our method and the exact results.
Notably, even in the regions where the spectrum becomes increasingly ``crowded'', the obtained eigenvalues are indistinguishable from those resulting from ED. 
It should be noted that if the eigenvalues are not far apart, the method will require more time to be able to distinguish between them. In Fig.~\ref{fig:dimer_spectrum_comparison} (c) we quantify the remarkable correspondence of the eigenvalues by showing that our method determines the $\lambda_j^{\rm eff}$ almost up to the numerical precision of ED.
This high-precision also characterizes the eigenmatrices, as shown in Fig.~\ref{fig:dimer_spectrum_comparison} (d). As a figure of merit, we plot the overlap between the eigenmatrices obtained through ED ($\eig{j}$) and those obtained with our method ($\eig{j}^{\rm eff}$), which again show a precision up to numerical error.

Another important question is how the Arnoldi-Lindblad time evolution fares against other possible methods to extrapolate the steady state from time evolution. Let us show here that our method is both faster and more precise, making it \textit{faster-than-the-clock}.
Consider, for example, $\langle\hat{n}_1\rangle = \langle\hat{a}_1^\dagger\hat{a}_1\rangle$, i.e., the expectation value of the particle number of the first cavity of the dimer.
In panel (a) of Fig.~\ref{fig:particlenumber} we show $\langle \hat{n}_1(t)\rangle$ for the same random initial state which we used for our algorithm. The vertical dotted lines indicate the final simulation times when our method reached convergence for the indicated number of eigenvalues and eigenmatrices [$\hat{\rho}^{(0)}$, $\hat{\rho}^{(4)}$ and $\hat{\rho}^{(9)}$]. This time is evidently much smaller than the one needed to reach the steady state through standard time evolution. 
Indeed, for $\gamma t=15$, $\langle \hat{n}_1(t)\rangle$ significantly differs from the steady-state solution $\langle\hat{a}^\dagger_1\hat{a}_1\rangle_{\rm ss}$, marked with a horizontal gray dashed line. 

Given the spectral structure in \eqref{eq:spectralsol}, one can extrapolate the steady-state expectation value from a long time dynamics.
For example, once transient fast processes have washed out, one has:
\begin{equation}\label{Eq:fit_curve}
    \langle \hat{n}_1(t)\rangle = \langle\hat{n}\rangle_{\rm ss} + \tilde{c}_1   e^{\lambda_1 t},
\end{equation}
where $\tilde{c}_1=c_1 \text{Tr}[\hat{a}_1^\dagger\hat{a}_1 \eig{1}]$, for $\eig{1}$ defined in \eqref{Eq:eigendecomposition} and \eqref{eq:spectralsol}.
From this, one can extrapolate $\langle\hat{n}\rangle_{\rm ss}$.
In panel (b) of Fig.~\ref{fig:particlenumber} we show as a function of time the relative error
\begin{equation}
    \|\delta \langle\hat{n}_1 \rangle\| /  \langle\hat{n}_1 \rangle_{\rm ss} = \| \langle\hat{n}_1 \rangle_{\rm ss}^{\rm eff} - \langle\hat{n}_1 \rangle_{\rm ss}\| /  \langle\hat{n}_1 \rangle_{\rm ss},
\end{equation}
where $\langle\hat{n}_1 \rangle_{\rm ss}^{\rm eff}$ has been obtained either through this fitting method (black solid line) or with our method for the case were convergency is reached for $\hat{\rho}^{(m)}$ for $m=0,\,4,\,9$ (horizontal lines). In other words, $\delta \langle\hat{n}_1 \rangle$ quantifies how accurate a prediction is at a time $\gamma t$.
It is clear that our method gives a result that is several orders of magnitude more precise than the one obtained with the fitting method, and a comparable precision can be obtained only for longer times.

Summing up, for this simple model, we have demonstrated that our method is efficient and remarkably predictive of the spectral structure of the Liouvillian superoperator. 

\subsubsection{The trimer}

We will now prove the efficiency of our method in a regime out of reach with Liouvillian ED by considering a trimer, i.e. 3 connected cavities. 
The parameter regime we consider is the same as in the dimer case (having rescaled $J$) and convergence is reached also in this case for a cutoff of $n_{max} = 7$. The Hilbert space now has a dimension $8^3 = 512$ and the whole Liouvillian has size $262144\times262144$.
Using our method, we can efficiently obtain the Liouvillian spectrum and eigenmatrices. In Fig.~\ref{fig:trimer} we show the 6 slowest processes in the Liouvillian spectrum of the trimer associated with $\hat{\rho}^{(5)}$.

\subsection{Asymmetric drive and time crystal in a dimer}\label{sec:assymetricdrive}

We now turn our attention to an asymmetrically driven Bose Hubbard dimer. Recently, it was shown that this system displays a time crystalline phase in the thermodynamic limit \cite{SeiboldPRA20}. 
In short, the onset of the time crystalline phase can be inferred from the scaling towards a well-defined thermodynamic limit, where everlasting oscillating phases emerge. 
Indeed, as one approaches the thermodynamic limit, the eigenvalues $\text{Re}\left(\lambda_{1,2}\right)\rightarrow 0$ but $\text{Im}\left(\lambda_{1,2}\right)\neq 0$, indicating the onset of a dissipative (or boundary) time crystal \cite{IeminiPRL18}.
As such, the phenomenology associated with time crystals and its spectral features strongly differ from those presented above, and the determination of the low-lying spectrum would require to fit the equations of motion using several parameters to capture the oscillating behaviors.
We refer the interested reader to Refs.~\cite{SeiboldPRA20,IeminiPRL18,MingantiarXiv20}.

\begin{figure}
    \centering
    \includegraphics[width=0.49 \textwidth]{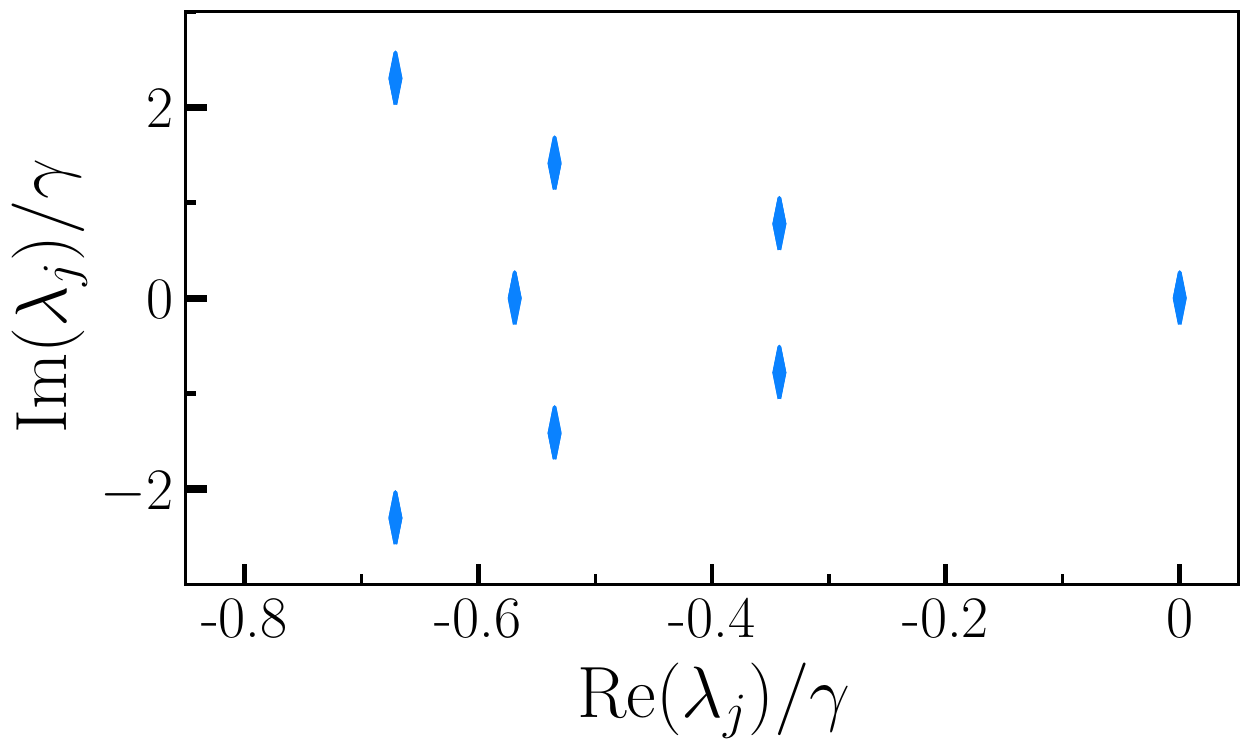}
    \caption{Results of $\hat{\rho}^{(8)}$ for a 
    Bose-Hubbard dimer with asymmetrical drive. The parameters are
    $F_1/\gamma= 8$, $F_2=0$, $U/\gamma=1/8$, $\Delta/\gamma=2$, and $J/\gamma=2$ as in Ref.~\cite{SeiboldPRA20}. The local Hilbert space has a cutoff at $n_{\rm max} = 27$.
    We obtained these results with a simulation of less than two hours.
     }
    \label{fig:dimerTC}
\end{figure}

Although time crystals can be observed only in the thermodynamic limit, for finite-size systems signatures of the time crystals can be argued from eigenvalues whose real part tends to zero while their imaginary part remains finite.
We show the low-lying spectrum of the asymmetrically driven DDBH in Fig. \ref{fig:dimerTC} for a parameter regime studied in  Ref.~\cite{SeiboldPRA20}. 
For these parameters, convergence is reached for $n_{\rm max} = 27$, meaning that the Hilbert space has dimension $784$ and $\LL$ is a $614656\times 614656$ matrix. Our results are in line with those of Ref.~\cite{SeiboldPRA20}, and in Fig.~\ref{fig:dimerTC} we observe eigenvalues $\text{Re}\left(\lambda_{1,2}\right)\approx -0.35\gamma$ and $\text{Im}\left(\lambda_{1,2}\right)\approx \pm 1$, whose associated eigenmatrices cause oscillations in the expectation values of the system for long times. 
These damped oscillations are the finite-size effects of a thermodynamic breaking of time translational invariance of the time crystal, much like the breaking of translational invariance in classical crystals.

\section{Floquet-Liouvillian systems}
\label{Sec:floquet}

\begin{figure}
    \centering
    \includegraphics[width=0.49 \textwidth]{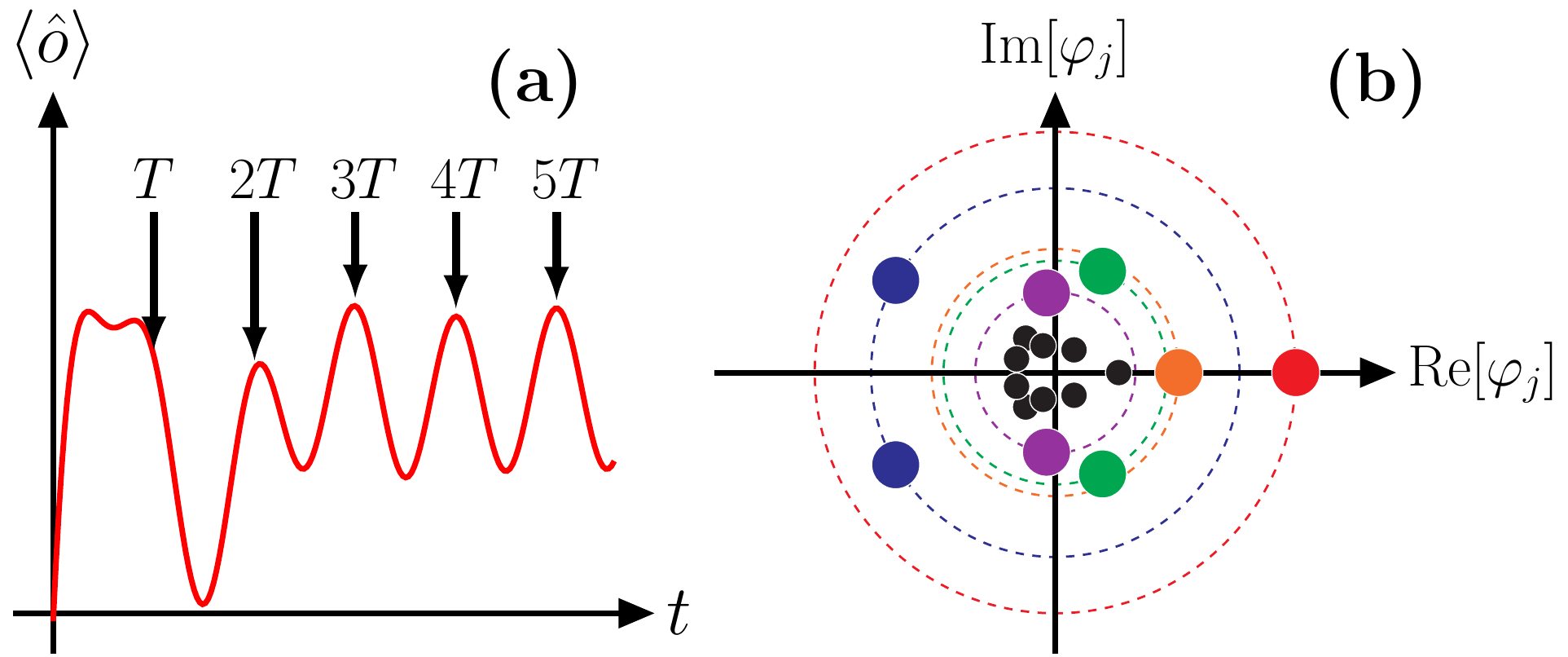}
    \caption{Time evolution of a time-dependent and periodic Liouvillian superoperator. (a) As in Fig.~\ref{fig:spectrum_extraction_alg}(a), but for a Liouvillian of period $T$. By monitoring the system every period $T$, the system converges to a stroboscopic stationary regime after an initial transient dynamics. (b) By introducing the Floquet map $\FF$, these characteristics are clear. Except from the stroboscopic steady state $\eig{0}$ characterized by $\varphi_0=1$, the eigenvalues characterizing the Floquet map eigenmatrices lie inside the unit circle, indicating decaying processes. Notice also the similarity between the spectrum of $\FF$ and that of the evolution operator of a time-independent Lindblad master equation $\EE$ in Fig.~\ref{fig:transformation_spectrum}.
    }
    \label{fig:spectrum_extraction_alg_FLOQUET}
\end{figure}

We now turn to periodically modulated quantum systems whose dynamics is described by a stroboscopic Lindblad master equation of period $T$, and whose time dependence cannot be gauged-out by an appropriate change of reference frame. 
The equation of motion of such a system is
\begin{equation}
\label{eq:Time_dependent_LME}
\partial_t \hat{\rho}(t) = \mathcal{L} (t) \hat{\rho}(t), \quad \mathcal{L}(t+T) = \mathcal{L}(t).
\end{equation}
The periodic temporal dependence of $\LL(t)$ can prevent the system from reaching a steady state.
Nevertheless, one can introduce a \textit{stroboscopic} steady state via the so-called \textit{Floquet evolution superoperator} (a Floquet map for Lindbladian systems).
Indeed, using the time ordering $\mathcal{T}$, we can formally solve \eqref{eq:Time_dependent_LME} as
\begin{equation}\label{eq:solution_Floquet}
\rhot = \mathcal{T} \left[\exp\left(\int_0^{t} \LL(t') dt'\right) \right] \hat{\rho}(0)= \FF(t, 0)\hat{\rho}(0),
\end{equation}
where  the evolution superoperator for the time-dependent Lindblad master equation $\FF(t, t_0)$ is the generalization of \eqref{Eq:Evolution_operator} to the time-dependent case.
Using the periodicity of the Liouvillian, we can always write
\begin{equation}
\label{eq:Floquet_Map}
\FF(t, 0) = \left[\FF(T, 0)  \right]^n \FF(t-n T, 0),
\end{equation}
for an appropriate integer $n$, where $\mathcal{F}(T, 0)$ is the Floquet map describing the evolution on a period $T$.
Furthermore, we can formally introduce the Floquet Liouvillian $\LL_{\rm F}$ as
\begin{equation}\label{Eq:Floquet_Liouvillian}
    \mathcal{F} \equiv \mathcal{F}(T, 0)  = \exp\left( \LL_{\rm F} T \right),
\end{equation}
where we drop the time dependence from $\mathcal{F}$.
The stroboscopic steady state is that state such that
\begin{equation}
   \LL^{\rm F} \sss^{\rm F} = 0, \quad {\rm or } \quad \mathcal{F} \sss^{\rm F} = \sss^{\rm F}.
\end{equation}
Similarly, we define the stroboscopic eigenspectrum of the Floquet system as [c.f. \eqref{Eq:correspondence}]
\begin{equation}
\FF \eig{j}^{\rm F} = \varphi_j^{\rm F} \eig{j}^{\rm F}=e^{\lambda_j^{\rm F} T}  \eig{j}^{\rm F}
\quad \Leftrightarrow \quad
\LL^{\rm F} \eig{j}^{\rm F} = \lambda_j^{\rm F} \eig{j}^{\rm F}.
\end{equation}

\subsection{Computing the Floquet map}
\label{Sec:Building_Map}
To obtain \eqref{eq:Floquet_Map} one can use the definition of the time ordering as
\begin{equation}
\mathcal{F} =\lim_{N \to \infty} \mathcal{T} \prod_{j=1}^{N} \left[\mathcal{I} + \frac{T}{N}\mathcal{L}\left(\frac{T \cdot j}{N}\right) \right],
\end{equation}
where $\mathcal{I}$ is the superoperator identity.
Although correct, it is extremely inefficient to compute $\mathcal{F}$ in this way. 

Different approaches use the so-called Magnus (or more involved) expansions \cite{BLANES_PhysRep2009, Blanes_IOP2010, Restrepo_PRL2016, LAPTYEVA_CompPhysComm2016, KUWAHARA2016,Sato_2020} to obtain an approximated form of $\mathcal{F}$.
The main idea behind this approximation is to rewrite 
\begin{equation}
    \LL^{\rm F}=\sum_{k=1}^{\infty} \LL^{\rm F}_{k},
\end{equation}
where, for instance,
\begin{equation}
\begin{aligned}\label{Eq:Low-order-Magnus}
\LL^{\rm F}_{1} &=\int_{0}^{T} \LL \left(t_{1}\right) \mathrm{d} t_{1}, \\
\LL^{\rm F}_{2} &=\frac{1}{2} \int_{0}^{T} \mathrm{~d} t_{1} \int_{0}^{t_{1}} \mathrm{~d} t_{2}\left[\LL \left(t_{1}\right), \LL \left(t_{2}\right)\right].
\end{aligned}
\end{equation}
Higher-order $\LL^{\rm F}_{k}$ require nested commutators, and we refer the interested reader to, e.g., Ref.~\cite{BLANES_PhysRep2009}.
Although viable, this method becomes quickly impractical for large systems, since it requires to compute the commutators of Liouvillian superoperators.
While in the case of a simple time dependence $\LL = \LL_0 + \LL_1(t)$, \eqref{Eq:Low-order-Magnus} requires to compute a few terms, for more involved time dependencies the Magnus expansion quickly becomes cumbersome.
Last but not least, at the end of the expansion one obtains $\LL^{\rm F}$, which still needs to be diagonalized.
As we discussed in Sec.~\ref{Sec:Methods}, this is not a viable strategy for large systems, because $\LL^{\rm F}$ suffers from the same problems as $\LL$.

A different approach computes the evolution superoperator using the time dynamics of a basis of the Liouvillian space \cite{Hartmann_NJP2017}. 
Indeed, let us consider 
\begin{equation}
\mathcal{F} \hat{\rho}_{i,j}, \quad \hat{\rho}_{i,j} = \ket{i}\bra{j}.
\end{equation}
Since $\hat{\rho}_{i,j}$ are an orthonormal basis of the operators space (i.e., any operator can be written as a linear combination of $\hat{\rho}_{i,j}$), we conclude that the matrix form of $\FF(T)$ can be obtained as
\begin{equation}\label{Eq:Construction_of_F}
\mathcal{F}_{[m=i \cdot (N+1) + j, :]} = \operatorname{vec}\left[  \hat{\rho}_{i,j}(T)\right],
\end{equation}
where $\mathcal{F}_{[m, :]}$ indicates the $m$th row of the evolution operator in its matrix form, and $\operatorname{vec}\left[  \hat{\rho}_{i,j}(T)\right]$ is the vectorized form of the initial density matrix $\hat{\rho}_{i,j}$ evolved for a time $T$.
The advantage of this method is double. 
On the one hand, it avoids the complexity of superoperator algebra.
On the other, the diagonalization of $\FF$ is much more efficient than that of $\LL^{\rm F}_{k}$, for the very same reason that makes the diagonalization of $\EE$ simpler than that of $\LL$ in the time-independent case.
Nevertheless, this method still needs to store a full Liouvillian superoperator and deals with diagonalization of large matrices.
Another heavy burden of this method is that to obtain $\FF$ one needs to evolve all the $\hat{\rho}_{i,j}$ for a time $T$.
For a Hilbert space of size $N$, this means an evolution of $N^2$ density matrices.

\subsection{Using Arnoldi-Lindblad time evolution}\label{sec:usingArnoldiLindblad}

\begin{figure*}[t]
    \centering
    \includegraphics[width=0.98 \textwidth]{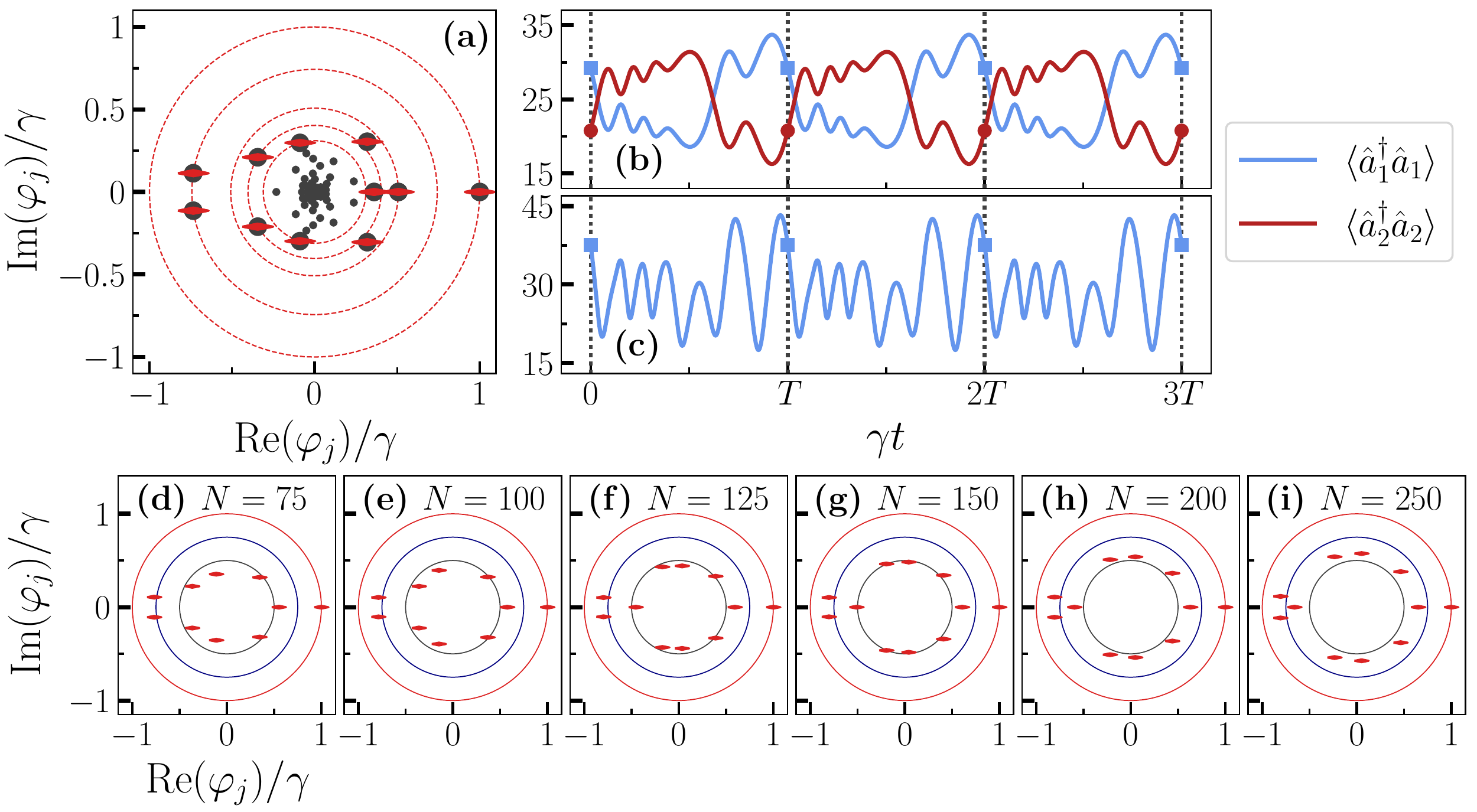}
    \caption{Results for the Bose-Hubbard dimer with a time dependent modulation presented in \eqref{eq:floquethamiltonian} and dissipation presented in \eqref{eq:floquetdissipation}. The chosen parameters, i.e., $UN/J = 1$, $f_0/J = 1$, $f_1/J = 3.4$, $\omega/J=1$, and $\gamma N/J = 0.2$, match those from Ref.~\cite{Hartmann_NJP2017} (to be consisted with our notation, a multiplication with a factor 2 was applied to $\gamma$). \textbf{(a)}, \textbf{(b)}, and \textbf{(c)} are for $N=50$. \textbf{(a)} Spectrum of the Floquet map $\FF$ obtained via exact diagonalization (gray dots) and our method for the 10 slowest states and their complex conjugates (red horizontal markers). The dashed lines represent the circles with radius equal to the absolute value of the eigenvalue $\varphi_j$. \textbf{(b)} The particle number expectation value $\langle \hat{a}_1^\dagger\hat{a}_1\rangle$ and $\langle \hat{a}_2^\dagger\hat{a}_2\rangle$, time evolved from the stroboscopic steady state $\sss^{F}\propto \eig{0}^{\rm F}$ obtained with our algorithm, i.e., the state whose eigenvalue is $\varphi_0 = 1$. \textbf{(c)} As a function of time, the real part of the particle number expectation value of the first cavity $\langle \hat{a}_1^\dagger\hat{a}_1\rangle$ (coinciding with that of the second cavity) for the slowest decaying state $\eig{1}^{\rm F}$ obtained with our algorithm, rescaled by the exponential damping $\exp(\vert\varphi_1\vert t)$. Given the degeneracy in the real part of $\varphi_1$ and $\varphi_2$, we assigned the value $\varphi_1$ to the one with negative imaginary part. Panels \textbf{(d)}-\textbf{(i)} the 10 largest eigenvalues (and their complex conjugates) of $\FF$ (red horizontal markers) are shown for \textbf{(d)} $N = 75$, \textbf{(e)} $N = 100$, \textbf{(f)} $N=125$, \textbf{(g)} $N=150$, \textbf{(h)} $N=200$, and \textbf{(i)} $N=250$. The red circle depicts the unit circle, the blue circle is the circle with radius $0.75$, and the gray circle has radius $0.5$.}
    \label{fig:Floquet_comparison}
\end{figure*}

The Floquet map possesses at least one (possibly degenerate) eigenvalue 1 and, similarly to the previous time-independent Liouvillian problem, all other eigenvalues lie inside the unit circle, see Fig.~\ref{fig:spectrum_extraction_alg_FLOQUET}(b). 
As such, we can use a variation of the Arnoldi-Lindblad algorithm to efficiently construct $\mathcal{F}$ using the system dynamics and simultaneously obtain its eigenvalues and eigenvectors allowing to reconstruct $\LL_{\rm F}$.
To do that, we simply need to check the system every period $T$, as sketched in Fig.~\ref{fig:spectrum_extraction_alg_FLOQUET}(a).
As in the previous case, this will construct the Krylov matrix of the system as
\begin{equation}\begin{split}
    K_n &= \left\{\hat{\rho}(0),\, \FF \hat{\rho}(0),\, \FF^2 \hat{\rho}(0)\, \dots ,\, \FF^n \hat{\rho}(0)  \right\}.
\end{split}
\end{equation}
From this, using the same procedure indicated in Sec.~\ref{Sec:Arnoldi-Lindblad}, one can construct the effective Floquet map $\FF_n^{\rm eff}$. 
In other words, this time the benefit is twofold:  
(i) the diagonalization of $\FF_n^{\rm eff}$ is much easier than the full diagonalization of $\FF$; (ii) $\FF_n^{\rm eff}$ is an excellent approximation of $\FF$.
However, our method becomes inefficient for large periods $T$, because time evolution itself becomes impractical.
Finally, we stress that our method can take into account nontrivial time dependence [e.g., Liouvillians of the form $\LL(t)=\sum_i\LL_i(t)$] without additional overhead. 
Such a task, instead, can be computationally very costly when considering Magnus-like expansions, as it is clear from \eqref{Eq:Low-order-Magnus}.

\subsubsection{Time-dependent Bose-Hubbard dimer}

To prove the efficiency and the validity of the method, we consider the Bose-Hubbard dimer which is governed by the time-dependent Hamiltonian
\begin{equation}
\begin{split}
    \hat{H}(t) &= \frac{U}{2}\sum_{j=1,2}\left(\crea{j}\right)^2  \left(\ann{j}\right)^2 - J\left(\hat{a}_1^\dagger\hat{a}_2 + \hat{a}_2^\dagger\hat{a}_1\right)\\
    & \quad \quad+ f(t)\left(\crea{2}\ann{2} - \crea{1}\ann{1}\right),
\end{split}
\label{eq:floquethamiltonian}
\end{equation}
where $f(t) = f_0 +f_1 \cos(\omega t)$ is a time dependent modulation of frequency $\omega$ which shifts the relative energies between the modes $1$ and $2$.
We consider that the corresponding Liouvillian is characterized by a dissipative process $\mathcal{D}\left[\hat{V}\right]$, where the jump operator $\hat{V}$ is given by 
\begin{equation}
    \hat{V} = \sqrt{\gamma}\left(\crea{1} + \crea{2}\right)\left(\ann{1} - \ann{2} \right),
    \label{eq:floquetdissipation}
\end{equation}
with $\gamma$ the dissipation rate.
The role of the dissipator can be easily understood in terms of the bonding [symmetric $\hat{c}=(\ann{1} + \ann{2})/\sqrt{2}$] and antibonding [antisymmetric $\hat{d}=(\ann{1} - \ann{2})/\sqrt{2}$] modes.
Indeed, $\hat{V} \propto \hat{c}^\dagger \hat{d} $ favors the bonding mode $\hat{c}$.
As such, the term $\hat{V}$ competes both with the static and the dynamics part of $\hat{H}(t)$ in determining the dynamics of the system.
These terms can be realized by a modulation of the system frequency, e.g., via a SQUID in circuit QED or by frequency shift by dispersive coupling in atomic physics. 
The jump operator, instead, is a ``collective'' dephasing which can emerge in atomic systems and could be engineered, for instance, by coupling a system to a noisy waveguide, or by averaging over stochastic Hamiltonian realizations.
Similar models have been extensively studied in the literature, both for Hamiltonian and dissipative cases, and we refer the interested reader to the discussions in Refs.~\cite{VardiPRL01,Trimborn_2008,WeissPRL08,PolettiPRL12, Hartmann_NJP2017}, and citations therein for systems characterized by similar processes.

This model is characterized by a strong $U(1)$ symmetry, meaning that both the Hamiltonian and the jump operator $\hat{V}$ commute with the total number of particles $(\hat{a}^\dagger_1 \hat{a}_1+ \hat{a}^\dagger_2 \hat{a}_2)$ \cite{BucaNPJ2012,AlbertPRA14,BaumgartnerJPA08}.
As such, the total number of particles $N=n_1(t)+n_2(t)$ is a conserved quantity [where $n_j(t)=\expec{\hat{a}^\dagger_j \hat{a}_j}(t)$], and the Liouvillian will admit multiple steady states.
However, we will set the initial state to be in the manifold containing $N$ excitations,
thus obtaining a well-defined Floquet map with a unique steady state for any finite $N$.

\subsubsection{Analysis of the Floquet map}

We proceed now to the analysis of the map $\FF$. At first we construct it with the method  described in \eqref{Eq:Construction_of_F} of Sec.~\ref{Sec:Building_Map} for $N=50$. Subsequently, we calculate the complete eigenspectrum of $\FF$ using exact diagonalization. This entire task requires approximately $300$s. The resulting eigenvalues $\varphi_j$ are shown as dark gray dots in panel (a) of Fig.~\ref{fig:Floquet_comparison}. We compute the 10 slowest $\varphi_j$ of $\FF$ with our method (red horizontal markers) in approximately $30$s. Once again, an excellent correspondence with the exact diagonalization is obtained.

Again, the obtained eigenvalues and eigenvectors are within numerical precision (not shown), and correctly describe the stroboscopic steady state and stroboscopic eigenmatrices.
For instance, in Fig.~\ref{fig:Floquet_comparison}(b) we plot the time evolution along several periods of $\expec{\hat{a}^\dagger_1 \hat{a}_1}$ (light blue curve) and $\expec{\hat{a}^\dagger_2 \hat{a}_2}$ (dark red curve) for a state initialized in $\sss^{\rm F}$.
The blue square and red dots represent the initial value, and are repeated at each period $T$, indicated by vertical dotted black lines.
Although $\sss^{\rm F}$ evolves, it periodically comes back to the initial value, as expected. 
Similarly, in Fig.~\ref{fig:Floquet_comparison}(c) we show that $\eig{1}^{\rm F}$ has a periodic structure if we take into account the exponential damping due to the eigenvalue $\varphi_1$. Notice also that $\eig{0}^{\rm F}\equiv \sss^{\rm F}$ represents out-of-phase oscillations, while $\eig{1}^{\rm F}$ represents in-phase oscillations of both cavities.

We investigate the longevity of the oscillating processes as we increase the size of the system in Figs.~\ref{fig:Floquet_comparison}(d-i). Despite an initial rearrangement of some eigenvalues [panels (d-f)], for large-enough $N$ the eigenvalues  begin to slowly approach the unit circle $\vert\varphi_j \vert=1$ [panels (g-i)].
The presence of these eigenvalues represent oscillating but not decaying processes because  $\varphi_j \neq1$. Thus, these are not associated with new stroboscopic steady states, but rather are associated with new oscillation frequencies in the system. 
This is a critical phenomenon, known as time crystallization, is known to emerge in dissipative systems \cite{Hartmann_NJP2017,IeminiPRL18}.

Concerning the efficiency of our method, we stress that for $N=250$ $\LL(t)$ and $\FF$ would have size $62500 \times 62500$, making it quite challenging to obtain $\FF$ and to subsequently diagonalize it, even on a supercomputer.
In contrast, our algorithm took just around $3000$s.

\section{Conclusions and Perspectives}
\label{Sec:Conclusion}

In this article, we have introduced a method to efficiently determine the steady state and the low-lying eigenvalues and eigenmatrices of the evolution operator of an open quantum system governed by a Lindblad master equation for (i) A time-independent Liouvillian; (ii) A Floquet (i.e., time-dependent but periodical) Liouvillian. 
Our method allows retrieving these features by a (shorter) time evolution of the open system, and consequently allows the calculation of the low-lying Liouvillian spectrum for system sizes that would be inaccessible through exact diagonalization. Although we focused on Runge-Kutta evolution of the system density matrix, our method works provided that we have access to the density matrices of the system throughout the time evolution.

We have tested our method on several examples of coupled bosonic nonlinear resonators, proving both its accuracy and efficiency.
We compared our method to other examples in literature, being able to retrieve and outperform the results of other articles (namely, Refs.~\cite{SeiboldPRA20,Hartmann_NJP2017}) using just our desktop PCs.
We are currently determining the spectrum of all-to-all connected problems, where the symmetries of open quantum systems allow for a further simplification of the computation, and where the original Liouvillian would be a $7\cdot 10^{10} \times 7\cdot 10^{10}$ matrix, and whose dynamics is difficult to extrapolate without the use of this method \cite{DolfTOAPPEARSOON, arnoldilindbladgithub}.

Even so, the presented algorithm can be optimized in order to ensure an even faster convergence.
For instance, the method may still present some difficulties for very large systems, since it may be necessary to store many density matrices before convergence is reached.
This problem could be avoided by using an Implicitly Restarted Arnoldi Method (IRAM). 
Such an optimization is beyond the purpose of this article, and the development of an IRAM time evolution is one of the future perspective of this work \cite{BaiBOOK, KressnerBOOK}, including deflation procedures to improve the convergence of the IRAM.
The presented method is model-independent, and it can be readily applied to any Lindblad master equation, provided that it is possible to numerically evolve it.
As such, we plan to release it as an open-source code, possibly to be integrated with the QuTiP libraries \cite{qutip1,qutip2}.

As we detailed along the main text, this method was developed because iterative methods, which normally are applied to determine the low-lying part of Hamiltonian matrices, tend to be ineffective for the Liouvillian problem, making it particularly challenging to determine the Liouvillian eigenspectrum.
Vice versa, this method does not show the same efficiency for Hamiltonian systems (i.e., no dissipation) since the eigenvalues of a Hamiltonian are purely real and therefore the Hamiltonian time evolution results in an operator whose eigenvalues all lie on the unit circle.
Consequently, there is no convergence towards a steady state (the wave functions do not decay as time evolves) and for large systems the unit circle becomes overcrowded with eigenvalues.

Concerning extensions of our method to other techniques to investigate open quantum systems, preliminary testing has shown that it should be possible to generalize it to the method of quantum trajectories. This is an exciting perspective, particularly concerning the statistical convergence of the density matrix reconstruction with the quantum trajectory method (one needs to average over an infinite number of quantum trajectories to obtain the exact evolution of the master equation \cite{DalibardPRL92,Molmer1993,DaleyAdvancesinPhysics2014}). 
Another future outlook is to extend these ideas to approximated time evolutions of open quantum systems, such as those stemming from truncated Wigner \cite{Foss-FeigPRA17,VicentiniPRA18,VicentiniPRA19,SeiboldPRA20} and Gaussian ansatzes \cite{VerstraelenPRA20,VerstraelenPREV20}. In these cases, the expectation values along a trajectory allow to approximately reconstruct the expectation values of all the operators. 
And since operators in the Heisenberg picture evolve via $\LL^\dagger$ and $\EE^\dagger$, by selecting an appropriate basis of operators we should be able to assess the spectral properties of an open quantum system even without knowing $\rhot$.

\acknowledgments{
The authors acknowledge useful discussions with Alberto Biella, Riccardo Rossi, Davide Rossini, Vincenzo Savona, David Schlegel, Dries Sels, and Michiel Wouters. 
D.H. is supported by UAntwerpen/DOCPRO/34878.}
\appendix

\section{Pseudo-codes}
\label{App:Pseudo-codes}
Here, we provide the pseudo-codes to perform Arnoldi iteration algorithm on a generic superoperator and the pseudo-code for the Krylov time evolution for the Lindblad master equation problem. 
The Arnoldi-Lindblad time evolution implemented in Python, with an example, can be found in Ref.~\cite{arnoldilindbladgithub}. Upon reasonable request, the Python code used in the article can be provided.

\begin{figure*}
\begin{minipage}{\linewidth}
\begin{algorithm}[H]
    \caption{Arnoldi iteration}
    \label{Alg:ArnoldiIteration}
    \begin{flushleft}
    \noindent \textbf{Input:} $\hat\sigma$, a random initial matrix; 
    $\LL$, the superoperator; the number $m$ of required eigenvalues; tolerance $\tau$.\\
    \noindent\textbf{Output:}  The $m$ pairs ($\lambda_j^{\rm eff}$, $\eig{j}^{\rm eff}$) of approximated eigenvalues and eigenmatrices.
    \end{flushleft}
    \begin{algorithmic}[1]
    \State $\hat{\sigma}_1\gets \hat\sigma/\left\|\hat\sigma \right\| $ \Comment{\{initial matrix\}}
    \State $k\gets 1$
    \While {Convergence is not reached}
        \State $k\gets k+1$
        \State $\hat{\nu} \gets \LL \hat{\sigma}_{k-1}$ \Comment{\{new and non-orthonormal element of the basis\}}
        \For{$1\leq j \leq k-1$} \Comment{\{Gram-Schmidt orthonormalization\}}
        \State $l_{j, k-1}\gets\operatorname{Tr}\left[\hat{\sigma}_j^\dagger \hat{\nu} \right]=\braket{\hat{\sigma}_j|\hat{\nu}}$ \Comment{\{recall that $l_{j, k-1}$ are the elements of $\LL_n^{\rm eff}$ in \eqref{Eq:Effective_Liouvillian}\}}
        \State $\hat{\nu}\gets \hat{\nu} - l_{j, k-1} \hat{\sigma}_j$
    \EndFor
    \State $l_{k, k-1} \gets \left\|\hat{\nu} \right\|$
    \State diagonalize $\LL_n^{\rm eff}$ and define the pairs ($\lambda_j^{\rm eff}$, $\eig{j}^{\rm eff}$)
    \If{$l_{k, k-1}=0$ (numerically) \text{ \textbf{or} } $\|(\LL-\lambda_l^{\rm eff}) \eig{l}^{\rm eff} \|<\tau$ \text{ for all $l\leq m$}}
        \State Convergence is reached
    \Else
        \State $\hat{\sigma}_k \gets \hat{\nu}/ \left\|\hat{\nu} \right\|$
    \EndIf
    \EndWhile
    \end{algorithmic}
\end{algorithm}
\begin{algorithm}[H]
    \caption{Arnoldi-Lindblad time evolution}
    \label{Alg:ArnoldiTimeEvolution}
    \begin{flushleft}
    \noindent \textbf{Input:} $\hat\rho(0)$, a random initial matrix; 
    $\hat{H}$ the Hamiltonian (possibly time-dependent); $\{\hat{J}_\mu\}$ the set of jump operators (possibly time-dependent); the time $T$ for each step (or the Floquet period); the number $m$ of required eigenvalues; tolerance $\tau$.\\
    \noindent\textbf{Output:}  The $m$ pairs ($\epsilon_j^{\rm eff}$, $\eig{j}^{\rm eff}$) of approximated eigenvalues and eigenmatrices of the $\EE$ (the Liouvillian evolution operator) or of $\FF$ (the Floquet map). From these, the (Floquet) Liouvillian eigenvalues can be readily obtained.
    \end{flushleft}
    \begin{algorithmic}[1]
    \State define the function $\EE$ (even in the Floquet case, where it should be $\FF$) which returns $\hat{\rho}(t+T)$ given $\hat{\rho}(t)$ using $\hat{H}$ and $\{\hat{J}_\mu\}$. We recall that $\LL$ or $\EE$ are never explicitly written in super-operator form
    \State $\hat{\sigma}_1\gets \hat{\rho}(0)/\left\|\hat{\rho}(0) \right\| $ \Comment{\{initial matrix\}}
    \State $k\gets 1$
    \While {Convergence is not reached}
        \State $k\gets k+1$
        \State $\hat{\nu} \gets \EE \hat{\sigma}_{k-1} $ \Comment{\{new and non-orthonormal element of the basis via time evolution of $\hat{\sigma}_{k-1}$ for $T$\}}
        \For{$1\leq j \leq k-1$} \Comment{\{Gram-Schmidt orthonormalization\}}
        \State $e_{j, k-1}\gets\operatorname{Tr}\left[\hat{\sigma}_j^\dagger \hat{\nu} \right]=\braket{\hat{\sigma}_j|\hat{\nu}}$ \Comment{\{recall that $e_{j, k-1}$ are the elements of $\EE_n^{\rm eff}$ in \eqref{Eq:Effective_Evolution}\}}
        \State $\hat{\nu}\gets \hat{\nu} - e_{j, k-1} \hat{\sigma}_j$
    \EndFor
    \State $e_{k, k-1} \gets \left\|\hat{\nu} \right\|$
    \State diagonalize $\EE_n^{\rm eff}$ and define the pairs ($\epsilon_j^{\rm eff}$, $\eig{j}^{\rm eff}$)
    \If{$e_{k, k-1}=0$ (numerically) \text{ \textbf{or} } $\|\EE \eig{l}^{\rm eff} - \epsilon_l^{\rm eff}\eig{l}^{\rm eff}\|<\tau$ \text{ for all $l\leq m$}}
        \State Convergence is reached
    \Else
        \State $\hat{\sigma}_k \gets \hat{\nu}/ \left\|\hat{\nu} \right\|$        
    \EndIf
    \EndWhile
    \end{algorithmic}
\end{algorithm}
\end{minipage}
\end{figure*}

\section{Iterative methods for the Liouvillian and their problems}
\label{App:Iterative}

A legitimate question is why standard applications of the Arnoldi method prove ineffective for the Liouvillian superoperator.
We briefly comment on them here, and we refer the reader to more details in the literature \cite{TrefethenBOOK}.

\subsection{Shifted Arnoldi}
\label{Sec:Shifted}
At first, we notice that in many Hamiltonian applications, where the spectrum is real, an efficient way to compute the Hamiltonian ground state is the shifted Arnoldi method.
Suppose that $\hat{H}$ is a $(N+1)\times(N+1)$ Hamiltonian whose ground state energy is $E_0$, while the largest energy is $E_{N}$.
If we were now to compute the spectrum of $\hat{C} = -\hat{H} + \mu  \hat{\mathds{1}}$ for $\mu>E_N/2$, and since $\hat{C}$ has the same eigenvectors as $\hat{H}$, the ground state would become the eigenvector whose eigenvalue is the largest.
In other words, a shifted Arnoldi method could quickly compute the ground state by finding a boundary for $\hat{H}$, and setting $\mu $ consequently.

The question is more tricky when discussing the Liouvillian.
Indeed, in the Liouvillian case, the eigenvalues are complex.
As such, this time we need to choose 
\begin{equation}
    \mu > \max_{j \neq 0} \frac{|\lambda_j|^2}{-2 \Re{\lambda_j}},
\end{equation} so that $\eig{0}$  becomes the eigenmatrix of the largest eigenvalue \textit{in magnitude} of the superoperator ${\mathcal{C} = \LL + \mu  \mathcal{I}}$.
Although mathematically well-defined, this approach can be inefficient for many physical systems, where $\mu$ becomes so large that the operator  $\mathcal{C}$ is numerically dominated by it.
This is the case for open interacting systems, because the imaginary part of $\lambda_j$ grows roughly quadratically with the cutoff (this is the effect of the interaction).
Hence, for a large-enough system, the imaginary part of $\lambda_j$ is the largest contribution to $\mu$, making it necessary to introduce very large $\mu$ in order to make for the steady state to be the eigenmatrix of the largest eigenvalue.

Although there are no general results on the convergence of Arnoldi iteration, in the best-case scenario the algorithm converges roughly geometrically with the distance between the eigenvalues \cite{TrefethenBOOK}.
Thus, to have a nondimensional equation, whose eigenvalues are within the unit circle, we consider $\mathcal{C}/\mu  = \LL/\mu  + \mathcal{I}$.
The steady state is the eigenmatrix of eigenvalue $1$ of $\mathcal{C}/\mu$, while all the other eigenvalues become $1+\lambda_j/\mu$. 
Thus, an upper bound for converge rate is given by $\max_{j\neq 0} |\lambda_0-\lambda_j|/\mu=|\lambda_j|/\mu$.
Such a number roughly scales as $\max_{j\neq 0} 2 |\Re{\lambda_j} / \Im{\lambda_j}| $ for a large-enough imaginary part of $\lambda_j$.
In other words, the convergence rate is given by the ratio between the smallest real part and the largest imaginary one.
Such a number can be minuscule.
For instance, for the Bose-Hubbard dimer we considered in Sec.~\ref{Sec:Dimer}, $\mu =  4.2 \times 10^{4} \gamma$.
As such, $\max_{j\neq 0} |\lambda_j|/\mu = 0.00033$, making the convergence of the Arnoldi iteration surely remarkably slow.
Indeed, we verified that for such a shifted Arnoldi method, the algorithm (i) Is very unstable and predicts the eigenspectrum with a high error;
(ii) In order to obtain the entire spectrum it requires to build the entire Liouvillian, resulting in no gain.

\subsection{Shifted-inverted Arnoldi}

Another option would be the use of the shift-inverted Arnoldi method.
Indeed, instead of solving  for the eigenspectrum of $\LL$, one could study the operator
\begin{equation}
    \mathcal{C} = (\LL - \zeta \mathcal{I})^{-1}.
\end{equation}
The eigenmatrices of $\mathcal{C}$ are the same as for the Liouvillian, but the eigenvalues would read
\begin{equation}
    \kappa_j = \frac{1}{\lambda_j - \zeta},
\end{equation}
making $\eig{0}$ the eigenmatrix of largest $\kappa_j$.
In this regard, the Arnoldi iteration could be applied by computing $\{\mathcal{C}^n \hat{\sigma}\}$ for a random state $\hat{\sigma}$ if we were to know how to write $\mathcal{C}$.

If, however, we do not know $\mathcal{C}$, as in most cases, a first option would be to invert it numerically (e.g., via LU decomposition).
Such a method, however, relies on the decomposition of the \textit{entire Liouvillian}, an operation which quickly becomes numerically untreatable.
As such, the computational and memory intensive portion of the method lies in the LU factorization of $\LL$.
Even for small examples as those considered in the main text, such a problem is also known to have several numerical instabilities, making the solution to the problem fairly difficult, and requiring a good preconditioning.

One can still try to iteratively solve this problem.
Indeed, by considering
\begin{equation}
    \mathcal{C} \hat{\sigma} = \hat{\chi}, 
\end{equation}
we can obtain by $\hat{\sigma}$ formally rewriting
\begin{equation}\label{Eq:Invesion_iterative}
    (\LL - \zeta \mathcal{I}) \hat{\chi}=\hat{\sigma}. 
\end{equation}
Since the Arnoldi iteration relies on the consequent application of $\mathcal{C} $ to $\hat{\sigma}$, one can avoid the explicit inversion by solving \eqref{Eq:Invesion_iterative} to obtain $\hat{\chi}$ given  $\hat{\sigma}$.
Such a task can be performed using iterative solvers.
In this way, we could, in principle, avoid the numerical cost of the application of the full Liouvillian, but we could restrict ourselves to the subsequent application of $\hat{H}$ and $\hat{J}_\mu$ on a density matrix.
Several remarks are, however, necessary.
First, we notice that the resolution of \eqref{Eq:Invesion_iterative} via iterative methods should be performed at each step of the Arnoldi iteration.
As such, the numerical cost of this procedure is comparable to that of a time evolution, where we continuously compute the action of $\hat{H}$ and $\hat{J}_\mu$ on $\rhot$ via, e.g., a Runge-Kutta algorithm.
Second, while iterative methods require less memory and fewer numerical operations than the exact LU decomposition, these methods usually require preconditioning to achieve a sufficient tolerance level and a reasonable convergence rate.
This problem is not only emerging in open quantum systems, but it is also known to affect Hamiltonian systems, as also discussed in Ref.~\cite{PietracaprinaSciPost18}.

Let us also remark that the exponential map considered in the main text tends to have none of the problems of the shifted and shifted-inverted Arnoldi method.
Indeed, the advantage of the exponential map is that the imaginary part of the eigenvalues $\lambda_j$ plays no role in determining the magnitude of the eigenvalues $\epsilon_j$.
As such, the exponential is not affected by the ``crowding'' of eigenvalues described earlier in Appendix~\ref{Sec:Shifted}. 
Furthermore, possible numerical instabilities in the construction of the Arnoldi basis are suppressed by the contractive nature of the exponential map.
In other words, one of the advantages of the method presented here is that, no matter the details of the physical system, if one is capable of computing the dynamics, it provides the low-lying eigenvalues and eigenvectors with high precision.


%

\end{document}